\documentclass{WileyMSP-template}

\usepackage[separate-uncertainty = true,multi-part-units=single]{siunitx}
\usepackage{xr}
\usepackage{cite}
\usepackage{newtxtext,newtxmath}
\DeclareMathOperator{\atantwo}{atan2}
\usepackage[T1]{fontenc}

\usepackage{placeins}
\usepackage{newfloat}
\DeclareFloatingEnvironment[name=Movie]{movie}

\begin{document}
\pagestyle{fancy}
\rhead{}
\title{Organic Semiconductor Alignment via Confinement in \newline Vapor-Guided Droplets}
\noindent\maketitle

\noindent\author{Robert Malinowski},\textsuperscript{1}
\author{Alessandro Rossi},\textsuperscript{2}
\author{Lewis M. Cowen},\textsuperscript{1}
\author{Peter A. Gilhooly-Finn},\textsuperscript{1}
\author{Michael A. Parkes},\textsuperscript{1}
\author{Ming-Hao Chang},\textsuperscript{3}
\author{Yu-Cheng Chiu},\textsuperscript{3}
\author{Ioannis Papakonstantinou},\textsuperscript{2}
\author{Matthew O. Blunt},\textsuperscript{1}
\author{Bob C. Schroeder},\textsuperscript{1}
\author{Giorgio Volpe\textsuperscript{1,*}}

\begin{affiliations}
\noindent\textsuperscript{1} Department of Chemistry, University College London, 20 Gordon Street, London WC1H 0AJ, United Kingdom.
\textsuperscript{2} Photonic Innovations Lab, Department of Electronic and Electrical Engineering, Torrington Place, WC1E 7JE, United Kingdom.
\textsuperscript{3} Department of Chemical Engineering, National Taiwan University of Science and Technology, Taipei City 106335, Taiwan. \\
\textsuperscript{*} Email Address: g.volpe@ucl.ac.uk
\end{affiliations}

%\noindent\keywords{Droplets, Organic Semiconductors, Printing, Molecular Self-assembly, OFETs}

\begin{abstract}

Organic semiconductors are lightweight, solution-processable materials with strong potential for printed and flexible electronics, from deformable displays to wearable sensors. Despite significant advances in materials synthesis and manufacturing, controlling molecular and mesoscale alignment during deposition remains a central challenge, as film morphology critically governs charge transport and device performance. Here, we demonstrate that flows developing within the intrinsically confined volume of microliter vapor-guided 
droplets can be harnessed to produce highly aligned organic semiconductor films. As droplets move in response to an external vapor source, internal flows align organic semiconducting nanowires within the droplet prior to deposition, yielding films with pronounced directional order. Organic field-effect transistors fabricated with this approach exhibit $\sim$40\% enhancement in saturation current  relative to spin-coated controls. Beyond improved device performance, the contactless and compact nature of our method enables the deposition and alignment of organic semiconductors on curved and flexible surfaces. More broadly, vapor-guided droplets offer a scalable framework for the confinement-induced alignment of functional soft materials, with potential for integration into existing additive manufacturing platforms for flexible electronics and beyond.
\end{abstract}

\section*{Introduction}

Organic semiconductors (OSCs) – carbon-based materials exhibiting semiconducting properties – offer unique opportunities for lightweight, flexible, solution‑processable and potentially sustainable electronics, enabling emerging technologies such as deformable displays and wearable sensors \cite{bronstein2020role, han2022materials}. 
While the highly crystalline structures of some inorganic materials offer efficient charge transport \cite{wang2013structure,khim2018uniaxial}, organic semiconductor films often exhibit structural disorder, polymorphism, and anisotropic charge transport \cite{khim2018uniaxial}.
Despite significant advances in molecular design and synthesis \cite{chen2023recent,zhang2025recent}, achieving a high level of alignment in printed OSC films remains a critical bottleneck. To date, the performance of organic electronic devices remains fundamentally hampered by the difficulty of controlling molecular and mesoscale alignment during solution processing \cite{khim2018uniaxial,wang2013structure}. 

Beyond off-center spin-coating\cite{chaudhary2019highly}, conventional continuous deposition methods – including blade coating \cite{chu2016toward,persson2017nucleation,wang2024green
}, meniscus‑guided coating\cite{delongchamp2009controlling,reinspach2016tuning,chang2016alignment,shaw2016direct,jo2018large,khim2013Simple,lee2021solutal}, and dip‑coating\cite{xue2010formation,wang2012microstructure,kim2019uniform,tao2025polymer
} – can impose directional order through controlled flow \cite{tao2025polymer}, capillary forces\cite{chu2016toward,persson2017nucleation,delongchamp2009controlling,reinspach2016tuning,chang2016alignment,shaw2016direct,jo2018large,khim2013Simple,xue2010formation,wang2012microstructure}, external electric fields\cite{li2026programmable} and even surface tension gradients \cite{wang2024green,lee2021solutal}. Additional strategies can apply shear via an immiscible liquid substrate, either through droplet spreading \cite{pandey2017solvent}, compression of floating inks\cite{ito2020100}, or rotation of supported films\cite{fujioka2025circular}. However, these approaches typically rely on mechanical contact {\cite{chu2016toward, persson2017nucleation, wang2024green, delongchamp2009controlling, reinspach2016tuning,chang2016alignment,shaw2016direct, jo2018large, khim2013Simple, lee2021solutal, wang2020mixed, zhang2017marangoni,diao2013solution, xue2010formation, wang2012microstructure,li2026programmable}, continuous liquid films, or engineered substrates \cite{xue2010formation,chen2022direct,pandey2017solvent,ito2020100,fujioka2025circular}, often yielding unidirectional order control and limiting their applicability to flexible, curved, or topographically complex surfaces\cite{wu2020fabrication}. 
Flow‑driven alignment is also exploited in  techniques such as wet‑spinning\cite{zhang2008macroscopic,
jalili2011one,tong2025super}, electrospinning\cite{richard2013molecular}, and extrusion{\cite{lee2017large,kotikian20183d,zhou2019perovskite,lu2025high}, although limited to fiber formation.
In contrast, droplet‑based printing technologies (e.g., ink-jet printing) offer spatially programmable patterning precision, material efficiency, and compatibility with additive manufacturing\cite{lukyanov2024inkjet}. Yet, discrete droplets typically suffer from uneven deposition due to evaporation and pinning artifacts \cite{deegan1997capillary,he2023binary}. Moreover, deposits from individual droplets exhibit radial order at best \cite{hu2020general}, as these droplets inherently lack the continuous directional flow fields required to impose long-range order without resorting to surface modification \cite{liu2024improving} or embedding matrices \cite{rodlmeier2017controlled}. 

Here, we introduce a contactless strategy that harnesses the intrinsically confined internal flows of microliter vapor‑guided droplets to direct the alignment of organic semiconductors prior to deposition. As a model semiconductor, we work with poly(3-hexylthiophene) (P3HT), a reference semicrystalline conjugated polymer that, under appropriate conditions, self-assembles into semiconducting nanowires with a high degree of internal order \cite{wirix2014three}. Exposure to a remote external vapor source induces motion of binary droplets containing these P3HT nanowires and generates robust Marangoni flows that orient them within the confined volume of the droplet ({\bf Figure \ref{fig:1}}). Motion of these droplets across a surface leads to deposition of continuous films with pronounced mesoscopic preferential alignment. Due to this enhanced morphological control, organic field-effect transistors (OFETs) fabricated with this approach exhibit $\sim40\%$ enhancement in saturation current relative to spin-coated controls. Our method’s contactless actuation further allows deposition on curved and inclined surfaces (up to a 45\textdegree\,angle), highlighting its suitability for flexible and non‑planar electronics. 

\section*{Printing organic semiconductors with vapor-guided droplets}

To print organic semiconductors, a droplet must meet two key requirements simultaneously: it must move with minimal resistance and effectively dissolve the semiconductor material. Evaporating binary droplets (evaporating droplets composed of two volatile liquids) are an excellent candidate for this purpose, as they can move freely due to negligible contact angle hysteresis\cite{huethorst1991motion} when one liquid component displays a lower boiling point ($T_{\rm b}$) and higher surface tension ($\gamma$) than the other component \cite{pesach1987marangoni,cira2015vapour,malinowski2020nonmonotonic}. These material properties (along with a faster evaporation rate of the more volatile component at the edge of the droplet \cite{deegan1997capillary}) give rise to gradients of surface tension at the free interface of the droplet that generate and sustain recirculating Marangoni flows within it \cite{konvalina2015printing,homede2016printing,karpitschka2017marangoni,malinowski2018dynamic,malinowski2020nonmonotonic}. Typically, in an unperturbed droplet, these flows are radially symmetric, preventing spreading and reducing contact angle hysteresis \cite{konvalina2015printing,homede2016printing,karpitschka2017marangoni,malinowski2020nonmonotonic}. In the presence of an external vapor source, this symmetry can be readily altered, causing the droplet to move in response to the source (Figure \ref{fig:1}a) \cite{cira2015vapour,malinowski2020nonmonotonic}. Chlorobenzene (PhCl, $T_{\rm b}$ = 132°C
\cite{liu1994saturation}, $\gamma$ = 33 \unit{\milli\newton\per\meter} \cite{tahery2017surface}) and octamethylcyclotetrasiloxane (D4, $T_{\rm b}$ = 176°C \cite{wilcock1946vapor}, $\gamma$ = 19 \unit{\milli\newton\per\meter} \cite{myers1969surface}) are two miscible liquids with the required specifications for droplet motion that are also a good solvent pair for P3HT (20 wt\% D4/PhCl, Figure \ref{fig:1}a, Experimental Section): chlorobenzene (the major component in the droplet) is a solvent with a high solubility for this polymer that is commonly used for its solution processing \cite{xue2011different,machui2012determination}, while
octamethylcyclotetrasiloxane (the minor component in the droplet) is a liquid fully miscible with chlorobenzene that promotes P3HT nanowire formation (Experimental Section, {\bf Figure \ref{fig:S_UV}}). Moreover, both liquids are volatile enough to fully evaporate after deposition (due to their relatively low $T_{\rm b}$), thus leaving pristine films of P3HT behind (Figure \ref{fig:1}, {\bf Movie \ref{Movie:Motion}}). 

We deposited small D4/PhCl droplets (typical volumes $V_{\rm d} $ = 100 -- 300 \unit{\nano \litre}) containing P3HT nanowires (1 mg g\textsuperscript{-1} P3HT in 20 wt\% D4/PhCl) on solid substrates (Figure \ref{fig:1}, Experimental Section). On glass, these droplets feature a low contact angle ($\theta_{\rm c}\approx 14\degree$, Figure \ref{fig:1}b), thus increasing the contact area for deposition (typical droplet radius $R_{\rm D} \approx \qty{1.1}{\mm}$).
To move them and deposit P3HT nanowire films (Movie \ref{Movie:Motion}), we employed an external vapor source implemented by filling a glass capillary (inner radius $R_{\rm C}$ = 0.75 \unit{\milli\meter}, height $h_{\rm C}$ = 0.4 \unit{\milli\meter} from the substrate, Figure \ref{fig:1}b) with chlorobenzene (the major component of the binary droplet) and placing it off-center with respect to the droplet (Figure \ref{fig:1}a-b, Experimental Section). The infrared image in Figure \ref{fig:1}c shows the effect of the vapor source on droplet evaporation: the edge closer to the source experiences a reduced evaporation rate (lower evaporative cooling) with respect to the opposite side, as the surrounding vapor is richer in chlorobenzene diffusing from the capillary toward the droplet. As a consequence, the gradient of surface tension at the droplet's free interface and the corresponding Marangoni flows within become radially asymmetric weakening at the closer edge and setting the droplet in motion toward the vapor source (Movie \ref{Movie:Motion}) \cite{malinowski2020nonmonotonic}. The droplet can then be guided over large distances ($\approx$ 2 cm, about ten droplet body lengths) before depositing all dispersed P3HT by continuously moving the vapor source at constant speed (here 250 \unit{\um\per\second}) \cite{malinowski2020nonmonotonic}.

On moving, these droplets deposit P3HT nanowires as a thin film of the same width as the droplet diameter (Figure \ref{fig:1}d, Movie \ref{Movie:Motion}). Polarized UV-Vis spectra confirm successful deposition, with the presence of characteristic 0-0 (604 \unit{\nm}), 0-1 (554 \unit{\nm}) and 0-2 (550 \unit{\nm}) electronic transition peaks (Figure \ref{fig:1}d), in good agreement with previous observations \cite{niles2012j,dou2020controlling}.  
We found, light polarized perpendicular to the droplet's motion exhibits a higher absorbance and a sharper 0-0 transition than light polarized parallel to it. This absorption anisotropy suggests that the P3HT polymer chains within the nanowires are preferentially aligned perpendicular to the droplet's motion direction, since light polarized along the polymer backbone (and, hence, along its transition dipole) is absorbed more strongly \cite{nagamatsu2003backbone,pace2019intrinsically}. Furthermore, the pronounced and sharp 0-0 transition observed in the perpendicular polarization is indicative of stronger coherence between exciton dipoles along the polymer chain, characteristic of J-aggregate formation, which is favorable for charge carrier mobility \cite{dou2020controlling}. This is in contrast to the 0-1 transition, which reflects stronger interchain excitonic interactions instead and is associated with H-aggregates with poorer charge transport properties \cite{dou2020controlling}. The comparatively weaker 0–0 peak in the parallel polarization thus indicates a reduced population of J-aggregates in this direction. Since P3HT polymer chains are known to orient across the width of the nanowires rather than along their length \cite{samitsu2008effective,niles2012j}, these UV-Vis spectra suggest that nanowires are preferentially aligned along the printing direction, with a well-ordered polymer conformation.

 \begin{figure}[h!]
	\centering
    \includegraphics[width=12cm]{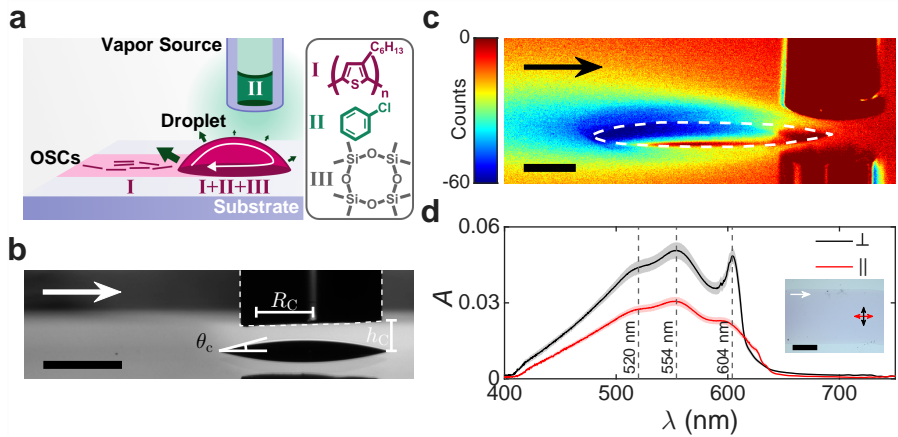}
	\caption{ \textbf{Printing organic semiconductors with vapor-guided droplets.} 
		(\textbf{a}) Schematic of a droplet depositing organic semiconductors (OSCs) on a substrate while moving in response to an external vapor source. Droplets are composed of ({\bf I}) P3HT nanowires and two volatile liquids, ({\bf II}) chlorobenzene (PhCl, also in the vapor source) and ({\bf III}) octamethylcyclotetrasiloxane (D4) (chemical formulae in the legend). The vapor source biases PhCl evaporative flux (green arrows) at the droplet's free interface, causing recirculating Marangoni flows (white arrow, simplification) and droplet's motion.    
        (\textbf{b}) Snapshot of a droplet (volume $V_{\rm d} = \qty{300}{\nano\litre} $, 20 wt\% D4/PhCl, 1 mg g\textsuperscript{-1} P3HT) with contact angle $\theta_{\rm c} \approx$ \ang{14} moving at a speed of 250 \unit{\um\per\second} beneath a capillary (white dashed lines) filled with PhCl of inner radius $R_{\rm C}$ = 0.75 \unit{\mm} at height $h_{\rm C}$ = 0.4 \unit{\mm} from the substrate.
		(\textbf{c}) Thermogram (in counts after background subtraction, Experimental Section) of a droplet (white dashed line, $V_{\rm d} = \qty{2}{\micro\litre}$, 20 wt\% D4/PhCl, no P3HT) moving at the same speed of 250 \unit{\um\per\second} following a capillary (red on the right-hand side). Evaporative cooling (lower counts) is visible on the droplet's side farthest from the capillary. 
        (\textbf{d}) Polarized UV-Vis absorption spectra (absorbance, $A$) from a 220 \unit{\um\squared} circular spot at the center of a printed P3HT nanowire film (inset, Movie \ref{Movie:Motion}) for light polarized perpendicular ($\perp$, black) and parallel ($||$, red ) to the printing direction (Experimental Section). Peak positions (vertical dashed lines) are in good agreement with those of P3HT nanowires deposited from toluene \cite{niles2012j}. Greater absorbance for the perpendicular polarization suggests alignment of polymer chains perpendicular to the printing direction \cite{nagamatsu2003backbone,pace2019intrinsically}, hence nanowire alignment parallel to it \cite{samitsu2008effective,niles2012j}. Shaded areas represent standard errors out of four independent measurements on different spots of the same film. All horizontal arrows show droplet motion direction within panel. All scale bars: 1 \unit{\mm}.
 }
	\label{fig:1}
\end{figure}

\section*{Role of internal droplet flows in aligning P3HT nanowires}

To elucidate the mechanism behind the alignment of the P3HT nanowires, we first need to better understand the flows that develop within the confined volume of a moving droplet ({\bf Figure \ref{fig:2}} and {\bf Figure \ref{fig:flows}}). For this purpose, we fed fluorescent dyes from both the droplet's front (Figures \ref{fig:2}a and \ref{fig:flows}a-b, {\bf Movie \ref{Movie:FlowFront}}) and its rear end (Figure \ref{fig:flows}c, {\bf Movie \ref{Movie:FlowBack}}) to visualize mass transport over time and extrapolate the internal three-dimensional flow (Figures \ref{fig:2}b and \ref{fig:flows}d, Experimental Section). These experiments reveal a complex internal flow field (Figure \ref{fig:2}b}, Figure \ref{fig:flows}) that resembles a radially asymmetric toroidal flow, in contrast to the symmetric toroidal flow observed during droplet evaporation in a uniform vapor field \cite{cira2015vapour,malinowski2020nonmonotonic}. The toroidal pattern is more compressed on the droplet's side closer to the vapor source. At the droplet's front (i.e., closest to the vapor source), a vortex forms that spirals outward along the droplet's edge before turning inward toward the droplet center (Figures \ref{fig:2}a-b and \ref{fig:flows}a-b,d, Movie \ref{Movie:FlowFront}). This spiral flow meets a recirculating rear flow that moves first along the free interface past the center of the droplet and then along the substrate toward its back (i.e., from regions of lower surface tension near the contact line toward regions of higher surface tension closer to the vapor source \cite{cira2015vapour,karpitschka2017marangoni, malinowski2020nonmonotonic}) (Figures \ref{fig:2}b and \ref{fig:flows}c-d, Movie \ref{Movie:FlowBack}).

As the flows recirculate within the confined volume of the droplet, they can align dispersed material with the direction of the flow \cite{li2006self,tao2025polymer,malinowski2020nonmonotonic}. For the P3HT nanowires, this flow-induced alignment is evident by observing the moving droplet under cross-polarization microscopy (Figure \ref{fig:2}c, Experimental Section): the droplet appears significantly brighter when illuminated with plane-polarized light at 45\textdegree\,with respect to the direction of droplet motion than when illuminated with plane-polarized light parallel to it, showing significant molecular anisotropy within the droplet's volume even before any material has been deposited. 
Light transmission at 45° to the droplet motion indicates polarization rotation arising from molecular alignment that is neither parallel nor perpendicular to the polarizer–analyzer axes. This implies that the optical axis of the nanowires is oriented either parallel or perpendicular to the droplet motion direction. Taken together with the observation in Figure \ref{fig:1}d, we can however infer that the nanowires must be preferentially aligned parallel to the droplet motion.
Moreover, the fact that the flows along the substrate at the rear of the droplet point toward the contact line is also consistent with the P3HT nanowires being aligned parallel to the direction of motion, as already observed in Figure \ref{fig:1}d.
The AFM analysis of the deposited P3HT films confirms the deposition of a dense network of nanowires (approximately one-to-two-nanowire thick) with a preferential alignment in the direction of the droplet motion in the center of the print (Figure \ref{fig:2}d and {\bf Figure \ref{fig:S_AFM_droplet}}a, Experimental Section). This preferential alignment is missing in large-area control films produced by spin-coating on glass (Figure \ref{fig:2}e and {\bf Figure \ref{fig:S_AFM_spin-coating}}a), a more traditional deposition technique. For droplet printed films, the extracted orientations of the nanowires show significant alignment with droplet motion (Figures \ref{fig:2}f and \ref{fig:S_AFM_droplet}b) and a random network for spin-coated films (Figures  \ref{fig:2}g and \ref{fig:S_AFM_spin-coating}b). This is confirmed by the nanowire orientation distributions displaying a narrow peak centered in the direction of motion for films printed with droplets (Figures \ref{fig:2}h and \ref{fig:S_AFM_droplet}b; mean angle from horizontal axis $\bar\varphi = \qty{0.01 \pm 0.07}{\radian}$; circular standard deviation $\sigma_{\rm s} = \qty{0.57\pm0.05}{\radian}$) and appearing uniform for the spin-coated samples (Figures \ref{fig:2}i and \ref{fig:S_AFM_spin-coating}b; $\bar\varphi = \qty{0.37\pm 0.54}{\radian} $; $\sigma_{\rm s} = \qty{1.05\pm0.13}{\radian}$) (Experimental Section). 

\begin{figure}[h]
	\centering
	\includegraphics[width=10cm]{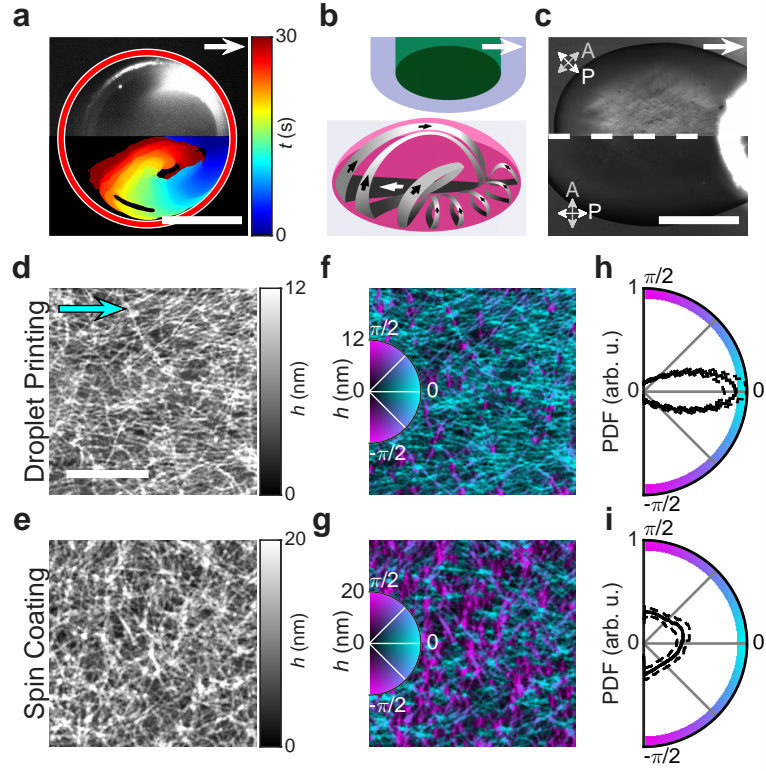} 
	\caption{
        \textbf{Alignment of P3HT nanowires by internal droplet flows.} 
        (\textbf{a}) Flow visualization by dye transport in a vapor-guided droplet (Movie \ref{Movie:FlowFront}, Experimental Section): exemplary fluorescence image at time 15 s (top half) and corresponding time evolution (color scale) over 30 s (bottom half, Experimental Section).
        (\textbf{b}) Schematic representation of the internal flows in a vapor-guided droplet extrapolated from dye transport analysis (Movies \ref{Movie:FlowFront} and \ref{Movie:FlowBack}, Figure \ref{fig:flows}). Toroidal flows develop as in sessile droplets\cite{cira2015vapour}, but the rear vortex dominates due to the asymmetric evaporative flux (Figure \ref{fig:1}c).
        (\textbf{c}) Cross-polarization microscopy images of two droplets moving toward a vapor source, taken with light polarized at 45\textdegree \, (top half) or parallel (bottom half) to their motion (polarizer P, white double-headed arrows; analyzer A, gray double-headed arrows), confirming alignment of the P3HT nanowires within the droplet's volume.
        Printing direction: top white arrows; scale bars: 1 mm.
        (\textbf{d}-\textbf{e}) $2.5 \times \qty{2.5}{\um\squared}$ AFM images, (\textbf{f}-\textbf{g})  respective orientation analysis (Experimental Section) and (\textbf{h}-\textbf{i}) orientation probability distribution functions (PDF, black solid lines) of P3HT nanowires (\textbf{d}, \textbf{f}, \textbf{h}) printed by a vapor-guided droplet moving in the direction of the top-left arrow in \textbf{d} and (\textbf{e}, \textbf{g}, \textbf{i}) spin-coated on glass, showing preferential alignment (cyan, 0 rad) for droplet printing only. Scale bar: 1 \unit{\um}.
        The color in \textbf{f} and \textbf{g} represents the P3HT nanowires' orientation and its intensity their height $h$ from AFM images. In \textbf{h} and \textbf{i}, each PDF and respective standard deviation (dashed lines) are from 8 different $5 \times \qty{5}{\um\squared}$ AFM images (Figures \ref{fig:S_AFM_droplet} and \ref{fig:S_AFM_spin-coating})
        }
	\label{fig:2}
\end{figure}

\section*{Electronic properties of P3HT nanowire films}

With improved alignment providing more directed pathways for charge transport \cite{khim2018uniaxial,memon2022alignment}, we expect the electronic performance of devices manufactured using P3HT nanowire films deposited by vapor-guided droplets to depend on the direction of printing. To confirm this, we designed custom electrodes to print organic field-effect transistors (OFETs) that are either aligned (electrodes perpendicular to droplet motion, $\perp$) or transversal (electrodes parallel to droplet motion, $\parallel$) to the printing direction ($V_{\rm d} = 300$ \unit{\nano \litre}, 20 wt\% D4/PhCl, 1 mg g\textsuperscript{-1} P3HT,  $\theta_{\rm c} \approx 15\degree$), thus allowing for a direct comparison between the electronic performance of the two configurations on the same films ({\bf Figure \ref{fig:3}}a-b, {\bf Movie \ref{Movie:MotionSide}}, Experimental Section). The output characteristics of both electrode configurations (Figure \ref{fig:3}c and {\bf Figure \ref{fig:S_parallel}}a) are typical for p-type OFET devices.
These output curves exhibit a clear transition from a linear regime at low drain-source voltages $V_{\rm DS}$ toward current saturation at increasingly negative $V_{\rm DS}$. The corresponding transfer curves (Figures \ref{fig:3}d and \ref{fig:S_parallel}b) show a $V_{\rm G}$‑dependent on/off current ratio of $\sim 3$ orders of magnitude and a well-defined threshold voltage at approximately $22$ \unit\volt, below which $I_{\rm DS}$ is strongly suppressed. The non-zero $V_{\rm th}$ is due to  oxygen doping arising from the measurements being performed in air \cite{kehrer2012temporal}. 
Minimal hysteresis is observed between forward and backward gate‑voltage sweeps. As expected, OFET performance is strongly dependent on nanowire alignment with respect to the electrodes (Figures \ref{fig:3} and \ref{fig:S_parallel}). Devices with nanowires aligned perpendicular to the electrodes exhibit superior characteristics compared to both the parallel configuration and spin‑coated controls ({\bf Figure \ref{fig:S_SC}}). 
In particular, the perpendicular electrode configuration results in $\sim42 \%$ increase in $I_{\rm DS}$ ($I_{\rm DS} = \qty{9.3}{\micro\ampere}$ at $V_{\rm G}$ = -80 \unit{\volt} and $V_{\rm G}$ = -30 \unit{\volt})} relative to spin-coated controls (Figures \ref{fig:3}c, e and \ref{fig:S_SC}), highlighting the beneficial role of the preferential nanowire alignment induced by our droplet-based deposition method. This improvement is observed even though the spin-coated films are $\sim 50 \%$ thicker (average thickness: 9.6 $\pm$ 4.1 \unit{\nm}) than the droplet-printed films (6.1 $\pm$ 2.7 \unit{\nm}).
In contrast, devices with nanowires aligned parallel to the electrodes display a 25\% decrease in maximum current relative to the spin-coated controls (Figure \ref{fig:3}e), consistent with charge transport occurring predominantly across, and not along, the nanowires.

The enhanced electronic properties of our films are also reflected in the extracted field-effect mobilities $\mu$ (Figure \ref{fig:3}f, Experimental Section), with the perpendicular electrode configuration displaying $\mu$ values $\sim 60\%$ higher on average for the same $V_{\rm G}$ (maximum $\mu =  7.5\times10^{-3}$ cm\textsuperscript{2}V\textsuperscript{-1}s\textsuperscript{-1}) than those of the parallel configuration in the investigated $V_{\rm G}$ range. Interestingly, the mobility of the random spin-coated nanowires transitions with gate voltage between the two curves obtained from films printed by droplets. This intermediate behavior likely arises from a complex interplay between the  organization of the nanowires, their resistivity, and the resistance associated with nanowire-nanowire junctions.  
Even though fewer junctions between nanowires are available in the aligned nanowires than in the spin-coated random network \cite{fata2020effect}, charge transport can proceed predominantly along their ordered polymer cores, enabling carriers to travel longer distances with relatively low resistance compared to transport that relies on hopping across nanowire junctions (as in a random network). However, at higher negative $V_{\rm G}$, the effective number of energetically accessible percolation pathways in the random network increases, partially compensating for the lack of long‑range alignment and leading to mobilities comparable to those of the perpendicular electrode configuration. 
By contrast, the devices in the parallel electrode configuration do not show the same convergence at higher negative $V_{\rm G}$, due to both reduced junction connectivity and unfavorable nanowire orientation with respect to the source–drain direction. As a result, while the mobilities of the spin-coated devices and those printed in the perpendicular electrode configuration are similar at $V_{\rm G} = -30$ \unit{\volt}, the advantage of nanowire alignment in terms of $\mu$ is more pronounced at lower gate voltages, where an enhancement of $\approx 80\%$ is observed at $V_{\rm G} = 10$ \unit{\volt}. 

\begin{figure}[h!]
	\centering
	\includegraphics[width=8cm]{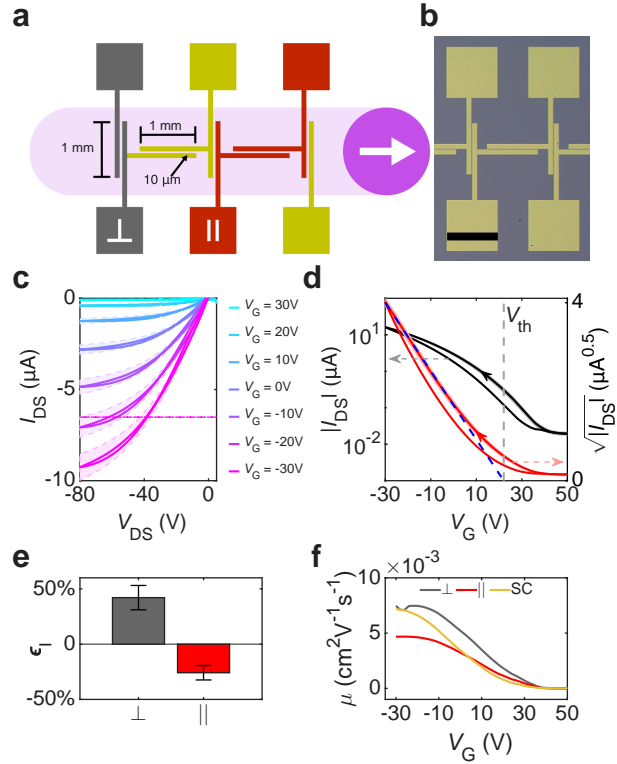} 
	\caption{
    \textbf{Printing organic field-effect transistors (OFETs) with vapor-guided droplets.}
        (\textbf{a}) Schematic and (\textbf{b}) microscope image of gold electrodes on an oxidized silicon wafer (Experimental Section). These were used to measure and compare electronic properties of P3HT films printed with vapor-guided droplets (this figure and Figure \ref{fig:S_parallel}, Movie \ref{Movie:MotionSide}) and spin-coated (Figure \ref{fig:S_SC}).
        They are arranged both perpendicular ($\perp$ gray) and parallel ($\parallel$, red) to the droplet's motion direction (white arrow), enabling measurements along either the printing direction or perpendicular to it. Inter-electrode channel length is and width are 1 \unit{\mm} and 10 \unit{\um}, respectively. Scale bar 1 \unit{\mm}.
        (\textbf{c}) Average p-type OFET output curves (out of eight measurements per curve, 4 independent chips, 2 devices each) showing drain-source current $I_{\rm DS}$ against applied voltage $V_{\rm DS}$ at different gate voltages $V_{\rm G}$ for P3HT nanowire films printed with vapor-guided droplets, measured in the $\perp$ electrode configuration in air. For reference, the horizontal dashed line shows the average $I_{\rm DS}$ at $V_{\rm G} = \qty{-30}{\volt}$ and $V_{\rm DS} = \qty{-80}{\volt}$ for 8 spin-coated devices (Figure \ref{fig:S_SC}). Output curves for the $\parallel$ electrode configuration and for spin-coated devices are shown in Figures \ref{fig:S_parallel} and \ref{fig:S_SC}, respectively.
        (\textbf{d}) Corresponding transfer curves showing $|I_{\rm DS}|$ (black) and $\sqrt{|I_{\rm DS}|}$ (red) as a function of $V_{\rm G}$ at $V_{\rm DS}$ = -80 \unit{\volt}. Threshold voltage $V_{\rm th} = \qty{22.1}{\volt}$ (dashed vertical line), extrapolated from fitting the linear region of the $\sqrt{|I_{\rm DS}|}$ curve (dashed blue line). Dashed arrows point to the vertical axis associated to each curve. In {\bf c} and {\bf d}, shaded areas show standard error on the forward scan (arrow heads in {\bf d}); backward scan errors are comparable but omitted for clarity.
        (\textbf{e}) Performance $\epsilon_{\rm I} = \big(I_{\rm max} / I^{\rm SC}_{\rm max} - 1 \big) \times 100\%$ (defined relative to spin-coated devices) of OFETs printed with vapor-guided droplets perpendicular ($\perp$) and parallel ($\parallel$) to the direction of motion, where $I_{\rm max}$ is the maximum measured current for either configuration and $I^{\rm SC}_{\rm max}$ that for spin-coated devices.
        Error bars show standard errors from 8 devices.
        (\textbf{f}) Hole mobility $\mu$ as a function of $V_{\rm G}$ in OFETs printed with vapor-guided droplets for the configurations perpendicular ($\perp$) and parallel ($\parallel$) to the direction of motion and for spin-coated devices (SC) (respectively calculated from the average transfer curves in {\bf d} and from Figures \ref{fig:S_parallel}b and \ref{fig:S_SC}b).
        }
	\label{fig:3}
\end{figure}

\section*{Printing OSCs on curved substrates with vapor-guided droplets}

\begin{figure}[h!]
	\centering
	\includegraphics[width=8cm]{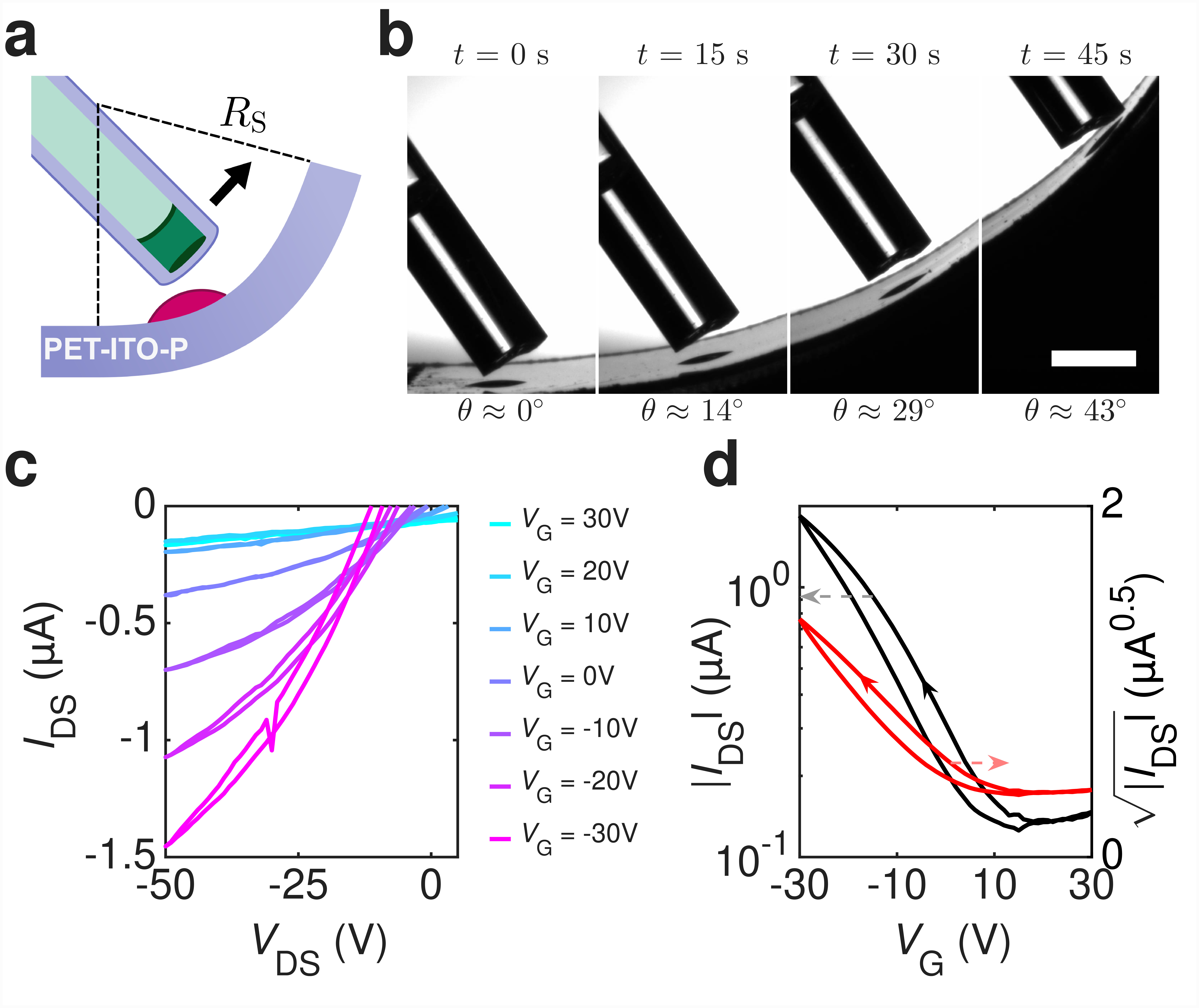} 
	\caption{
    \textbf{Printing OSC films and OFETs on curved substrates.}
        (\textbf{a}) Schematic and (\textbf{b}) time sequence of a vapor-guided droplet ($V_{\rm d} = 100$ \unit{\nano\litre}, 20 wt\% D4/PhCl, 1 mg g\textsuperscript{-1} P3HT) moving on a curved substrate (made of polyethylene terephthalate coated with indium tin oxide and paralyene, PET-ITO-P) with curvature radius $R_{\rm S} = 15$ \unit{\mm} patterned with flexible electrodes resembling the configuration in Figure \ref{fig:3}a-b (Figure \ref{fig:flexibleelectodes}, Movie \ref{Movie:Uphill}, Experimental Section). In \textbf{b}, the substrate inclination angle $\theta$ is indicated under each frame. Scale bar 2 \unit{\mm}.
        (\textbf{c}) Exemplary OFET output curves showing drain-source current $I_{\rm DS}$ against applied voltage $V_{\rm DS}$ at different gate voltages $V_{\rm G}$ for P3HT nanowire films printed with vapor-guided droplets on electrodes perpendicular to the droplet motion direction. Measurements are in air on the curved substrate.
        (\textbf{d}) Corresponding transfer curves showing $|I_{\rm DS}|$ (black) and $\sqrt{|I_{\rm DS}|}$ (red) as a function of $V_{\rm G}$ at $V_{\rm DS}$ = -50 \unit{\volt}. Arrow heads indicate forward scans and dashed arrows point to the vertical axis associated to each curve.
		}
	\label{fig:4}
\end{figure}

Beyond improved device performance, the contactless and compact nature of our method enables the
deposition and alignment of organic semiconductors on the curved surface of a flexible substrate ({\bf Figure \ref{fig:4}}a-b and {\bf Movie \ref{Movie:Uphill}}). More traditional methods such as spin-coating or continuous deposition approaches would instead struggle with this task due to their inherent limitations in accommodating non‑planar geometries while maintaining uniform film formation and alignment \cite{ding2016review}.
As can be seen in the time sequence of Figure \ref{fig:4}b and in Movie \ref{Movie:Uphill}, our vapor-guided droplets ($V_{\rm d} = \qty{100}{\nano\litre}$, 1 mg g\textsuperscript{-1} P3HT in 20 wt\% D4/ChPh, $\theta_{\rm c} \approx 15\degree$) can move uphill (up to an incline of $\theta \approx$ 45\textdegree) against gravity on a curved substrate (radius of curvature ($R_{\rm S}$ = 15 \unit{\mm}). The substrate was made of polyethylene terephthalate coated with
indium tin oxide and paralyene (PET-ITO-P) and patterned with gold electrodes configured as in Figure \ref{fig:3}a-b ({\bf Figure \ref{fig:flexibleelectodes}}, Experimental Section). The stage holding the capillary was programmed to move following the same curvature as the substrate. The forces at play on the droplet moving at tangential speed $v_{\rm D} = \qty{250}{\micro\meter\per\second}$ along the incline are the drive $F_{\rm vs}$ exerted by the vapor source, gravity $F_{\rm g} = -mg {\rm sin} \theta \approx -\qty{750}{\nano\newton}$ (with $m = \qty{108}{\micro\gram}$ the mass of the droplet and $g$ the acceleration due to gravity), and the kinetic viscous force $F_{\rm v} = - 3 \pi \eta \ell_{\rm n} R_{\rm D} \theta_{\rm c} v_{\rm D} \approx -\qty{5}{\nano\newton}$ due to the viscous stress at the base of the droplet on the solid surface ($\eta= \qty{1.05}{\milli\pascal\second}$ the dynamic viscosity and $\ell_{\rm n} \approx 14$ a dimensionless cutoff determined by the geometry of the droplet and the liquid constituting it \cite{malinowski2020nonmonotonic,malinowski2020advances}). From the force balance equation in the overdamped regime \cite{malinowski2020nonmonotonic,malinowski2020advances}, we can therefore estimate that the force $F_{\rm vs}$ that the vapor source can exert on the droplet is at least in the order of $F_{\rm vs} \approx - F_{\rm g} \sim \mathcal{O}(\qty{1}{\micro\newton})$.

The output and transfer characteristics of the devices printed on this curved substrate (Figure \ref{fig:4}c-d) confirm OFET behavior similar to the devices printed on a flat substrate (Figure \ref{fig:3}), albeit with a reduced output current ($I_{\rm DS} = \qty{-1.45}{\micro\ampere}$), $V_{\rm G}$-dependent on/off current ratio (one order of magnitude) and mobility ($\mu = 9.7\times10^{-4}$ cm\textsuperscript{2}V\textsuperscript{-1}s\textsuperscript{-1}). These differences likely result from limitations introduced through the in-house fabrication of our flexible substrates. These results nonetheless demonstrate the potential of vapor-guided droplets for printing working OSC devices on curved flexible substrates.

\section*{Conclusion}

In summary, we have demonstrated how the internal flows developing within the confined volume of a moving droplet guided by a remote external vapor source can be harnessed to align and print thin organic semiconductor films in a contactless way. Using nanowires of the reference OSC polymer P3HT, we demonstrated printing of working OFETs with enhanced electronic properties over spin-coated control films. By measuring the electronic properties of devices with electrodes perpendicular and parallel to the printing direction, we demonstrated that nanowire alignment within a device in a direction that is preferential to charge transport is critical to this improved electrical performance. This method can smoothly print functional devices on both flat and curved surfaces patterned with electrodes. This versatility highlights the potential for printing electronics beyond conventional two-dimensional wafer-based architectures, toward non-planar circuits that better conform to natural shapes, including the human body \cite{choi2021customizable}. The contactless nature of the approach further eliminates key challenges associated with many continuous deposition techniques, such as dip-coating, slot-die coating, and direct writing, the film properties of which can be highly sensitive to applicator's height and geometry \cite{ding2016review}.

Furthermore, printing with vapor-guided droplets uniquely bridges the gap between continuous meniscus-guided methods and discretized droplet-based techniques, most notably inkjet printing. It extends the precision of droplet deposition toward the formation of continuous, highly aligned films -- a critical requirement for fabricating high-performance organic polymer devices. In this context, our approach is well positioned to support the broader adoption of plastic electronics in applications poorly suited to silicon-based technologies, including flexible, non-planar, and space-constrained devices for applications in, e.g., wearable electronics \cite{choi2021customizable}.

Finally, this work opens a route toward the synthesis of self-patterning inks, in which vapor-guided droplets sense each other to interact and organize autonomously to generate complex structures \cite{cira2015vapour, thapa2025capillary}. When combined with additional Marangoni-driven phenomena, such as flower-like spreading \cite{wodlei2018marangoni} or timed spreading and contraction\cite{baumgartner2022marangoni}, this approach may enable autonomous pattern formation at sub-droplet length scales, as well as further morphological control through self-assembly \cite{forth2026towards}.

\section*{Experimental Section}

\threesubsection{Materials}
Regioregular poly(3-hexylthiophene-2,5-diyl) (P3HT) (Ossila, weight-averaged molecular mass {\it M}\textsubscript{W} = 60.2 kDa, number-averaged molecular mass {\it M}\textsubscript{n} = 28.7 kDa, dispersity Đ = 2.1, regioregularity RR = 98\%), chlorobenzene (PhCl) (Fisher Scientific, 99.8\%) and octamethylcyclotetrasiloxane (D4) (Fisher Scientific, 98\%) were used for droplet formulation. 
Methanol (Fisher Scientific, $\geq$99.8\%), acetone (Sigma-Aldrich, $\geq$99.5\%), hexane (Sigma-Aldrich, $\geq$95\%), and chloroform were used for P3HT purification. Sodium hydroxide (NaOH) (Fisher Scientific), isopropyl alcohol (Sigma-Aldrich, $\geq$99.5\%) and deionized water ($>$15 \unit{\mega\ohm}, obtained from an ELGA PURELAB® Chorus 2 water filtration system) were used for substrate cleaning. Glass slides (Epredia) were purchased from Fisher Scientific. Custom oxidised silicon wafers patterned with gold electrodes were purchased from ConScience AB and used for OFET measurements (Figure \ref{fig:3}). Flexible substrates (Figure \ref{fig:4}) were fabricated using polyethylene terephthalate (PET) films coated with indium tin oxide (ITO) (Thorlabs, 200 \unit{\um} thick) and Parylene-N coating (Speciality Coating Systems). Photoresist S1818 (MICROPOSIT Series, Kayaku Advanced Materials) and  developer MF-319 (Kayaku Advanced Materials) were used for patterning the gold electrodes onto these flexible substrates. Tetraphenyl-21\textit{H},23\textit{H}-porphine zinc and \textit{N},\textit{N}-Dimethylformamide were purchased from Aldrich and used for imaging flows within vapor-guided droplets (Figures \ref{fig:2} and \ref{fig:flows} and Movies \ref{Movie:FlowFront} and \ref{Movie:FlowBack}).

\threesubsection{P3HT Purification}
As-received P3HT was purified by sequential soxhlet extraction using methanol, acetone, hexane and chloroform for at least 6 hour each under nitrogen. P3HT is then precipitated from the final chloroform extraction with methanol and filtered. 
On purification, P3HT was analyzed by size exclusion chromatography (SEC) using a Shimadzu LC-2030c system with chlorobenzene as the eluent (80 °C, 1 \unit{\mL\per\minute}). Two mixed-bead polystyrene (PLgel MIXED-B, Agilent) columns connected in series were used as the stationary phase. Molecular weight averages ({\it M}\textsubscript{w} and {\it M}\textsubscript{n}) were determined by calibration against narrow polystyrene standards (EASICAL PS-1, Agilent) using an RID-40A detector.
P3HT analysis after purification showed {\it M}\textsubscript{W} = 59.5 kDa, {\it M}\textsubscript{n} = 41.9 kDa, Đ = 1.42, RR = 91\%. Purified P3HT was stored in a vacuum desiccator, protected from light exposure. 

\threesubsection{P3HT nanowire preparation}
Before each experiment, a 1 mg g\textsuperscript{-1} dispersion of P3HT nanowires (typical diameter $\approx 4$ nm) was prepared in a vial by dissolving 1 mg of P3HT in 1 g of 20 wt\% D4/PhCl stock solution (obtained by mixing 1 g of D4 with 4 g of PhCl). The vial was sealed and heated to 70°C in a water bath on a hot plate for 30 min, then left standing for 7 to 9 days, time needed for the P3HT nanowires to form fully (Figure \ref{fig:S_UV}a). We monitored P3HT nanowire growth with the peak at 610 nm in the UV-Vis absorption spectrum as it is the most separate from free molecular P3HT (Figure \ref{fig:S_UV}b). The dispersion was vortexed for 30 s before printing.

\threesubsection{Substrate cleaning}
Glass slides for the experiments in Figures \ref{fig:1} and \ref{fig:2} were cleaned at room temperature by sequentially sonicating in acetone (10 min), rinsing with DI water, sonicating in 1 M aqueous NaOH (10 min), rinsing with DI water until reaching neutral pH, sonicating in fresh acetone (10 min), and finally sonicating twice in DI water (10 min each). Patterned silicon wafers (Figure \ref{fig:3}) were cleaned by sequentially placing them in acetone twice, isopropyl alcohol, a solution of isopropyl alcohol and DI water (1:1 by volume) and DI water, each for 10 min at 40 °C.

\threesubsection{Setup for droplet printing and imaging}
All deposition experiments with droplets were performed and recorded using a custom-built printer under controlled conditions for temperature (21 $\pm$ 0.5)°C  and relative humidity (45 $\pm$ 5)\%. Substrates were placed on a motorized XY-stage enclosed within a custom perspex box to prevent air flow disturbances. Droplets were deposited on the substrate and guided using a glass capillary (inner radius $R_{\rm C}$ = 0.75 mm) containing a source of PhCl vapor (Figure \ref{fig:1}) and connected to a motorized XYZ-stage to control height $h_{\rm C}$ from and lateral displacement on the substrate. All deposition experiments were performed with $h_{\rm C}$ = 0.4 mm (Figure \ref{fig:1}a-b). 
Both values of $R_{\rm C}$ and $h_{\rm C}$ where chosen to achieve a strong attractive force of the vapor source on the droplet \cite{malinowski2020nonmonotonic}: $h_{\rm C}$ was set to bring the vapor source as close as possible to the droplet, while avoiding contact; $R_{\rm C}$ was chosen to create a strong compositional difference between the front and rear of the droplet, without saturating the vapor of the local environment and stopping chlorobenzene evaporation altogether. All stages could be programmed independently using a custom {\sc Matlab} application. For imaging in transmission, a white LED (Thorlabs, MNWHL4) was collimated using an achromatic lens (focal length $f$ = 30 mm) and used to illuminate the sample. This was then imaged using an objective lens ($f$ = 60 mm), an iris and a tube lens ($f$ = 75 mm) onto a color complementary metal-oxide semiconductor (CMOS) camera (Thorlabs, DCC1645C). Two linear polarizers (a polarizer and an analyzer) could be inserted before and after the sample at $90 ^{\circ}$ to each other for cross-polarization microscopy experiments (Figure \ref{fig:2}c). For imaging from the side, a cyan LED (Thorlabs, M505L4) was used to illuminate the sample, which was then imaged using an objective lens ($f$ = 100 mm), an iris and a tube lens ($f$ = 100 mm) onto a monochrome CMOS camera (Thorlabs, DCC1545M). All videos were recorded at 10 frames per second (f.p.s.). Printing started by placing a droplet (300 nL) on the substrate underneath the capillary using a glass syringe with repeater adapter (Trajan). The substrate was then moved relative to the capillary at a fixed speed, $v$ = 250 \unit{\um\per\second}, thus maintaining the droplet in frame when recording videos. Deposited films were then kept under vacuum for at least half an hour before electronic measurements.

\threesubsection{Thermography}
Thermographs of droplets on clean glass slides near a capillary (Figure \ref{fig:1}c) were taken using an FLIR A6703 MWIR camera at 60 Hz using an IR (3-5 µm) microscope lens (1x - 20 cm working distance, FLIR). A larger droplet volume ($V_{\rm d} = 2$  \unit{\micro\litre}) was used compared to printing experiments to better visualize thermal gradients. 

\threesubsection{UV-Vis spectroscopy}
UV-Vis spectra of P3HT nanowire dispersions were taken using a Shimadzu UV-3600i Plus UV-Vis-NIR spectrophotometer. Before measurement, a rectangular glass capillary (path length, 0.4 mm) sealed at one end was filled with the P3HT solution to a height of approximately 1 \unit{\cm}, tightly sealed with parafilm at its open end, and heated to 70 °C for 30 min, after which it was moved to room temperature. UV-Vis spectra were then measured at different times up to 11 days to observe nanowire formation (Figure \ref{fig:S_UV}). UV-Vis spectra (Figure \ref{fig:1}d) of deposited thin films were acquired using a custom optical setup instead. Samples were placed on the stage of a Leica DMI4000 B microscope coupled to a Kymera 328i-B2 spectrograph (150 lines/mm grating, blaze wavelength 500 nm) equipped with an Andor Newton Electron Multiplying Charge-Coupled Device (EMCCD) camera. The light from an incandescent bulb (RS components, 6.3 V, 200 mA) was collimated using an achromatic lens ($f$ = 30 mm) through a linear polarizer and directed through a 100 \unit{\um} pinhole. This aperture was imaged onto the sample using an objective (Olympus UPlanFL N, 4x, NA = 0.13), achieving a circular illumination spot of approximately 220 \unit{\um\squared}. The sample was then imaged using a second objective (Leica HC PL FLUOTAR, 20x, NA = 0.50) and a tube lens ($f$ = 40 mm) onto the slit of the spectrograph. A reference was taken through the glass substrate (averaged over three different spots 50 \unit{\um} apart) before measuring a spectrum at the center of the film (averaged over three different spots 50 \unit{\um} apart). The final spectra in Figure \ref{fig:1}d are an average of four such measurements along the respective print. 

\threesubsection{Flow visualization with fluorescent dyes}
A fluorescent ink for flow visualization (Figures \ref{fig:2} and \ref{fig:flows}, Movies \ref{Movie:FlowFront} and \ref{Movie:FlowBack}) was prepared by dissolving 5,10,15,20-Tetraphenyl-21\textit{H},23\textit{H}-porphine zinc (2 \unit{\mg}) in \textit{N},\textit{N}-Dimethylformamide (1 \unit{\g}) using sonication (10 min), followed by filtering through a glass fiber syringe filter (Whatman, 0.7 \unit{\um} pore size). A small amount of this ink was then printed on a clean glass slide using an ink-jet printhead (Microdrop Technologies, MD-K-140), dispensing one hundred $\approx$200 \unit{\pico\litre} droplets at the same location every $\approx$ 5 \unit{\s}. To introduce the dye from the front of the droplet (Movie \ref{Movie:FlowFront}), the capillary with the vapor source was placed 1.5 \unit{\mm} away from the deposited dye. A 300 \unit{\nano\litre} droplet composed of 20 wt\% D4/PhCl was placed under the capillary and led over the dye at 250 \unit{\um\per\s}. Similarly, to introduce the dye from the rear of the droplet (Movie \ref{Movie:FlowBack}), the capillary was placed 1 \unit{\mm} away from the deposit. A similar droplet was then placed under the capillary, which drew it over the dye; the droplet was then moved away from the deposit at the same speed. Videos of dye transport in droplets were recorded at 10 f.p.s. with our custom-built setup (\emph{Setup for droplet printing and imaging}) under blue light excitation (470 \unit{\nm}  LED, Thorlabs M470L5) with a long-pass filter (590 \unit{\mm} cut-on wavelength) in front of the acquisition camera. The recorded frames were processed to visualize dye mass transport over time. Each processed frame was obtained by averaging 20 raw frames and applying a Gaussian smoothing filter (kernel with a standard deviation of ten pixels), after which pixels outside the droplet profile were discarded. The effective frame rate was then reduced by a factor of three. Analysis then proceeded as follows: for each time point $t_n$ (with $n$ a non-negative integer), the difference $f_{n+1} - f_{n}$ between two consecutive processed frames was calculated; pixels with values above the 99.8\textsuperscript{th} percentile (99.5\textsuperscript{th} percentile for $n= 0$ only) were assigned to $t_n$ and excluded from further analysis. This results in a heat map where pixels are assigned the time at which they started contributing to the fluorescent signal meaningfully, thus highlighting mass transport within the droplet. In Figure \ref{fig:flows}c, the threshold was set at the 96\textsuperscript{th} percentile and applied at every step, due to faster flows and the smaller region of interest. 

\threesubsection{Spin-coating}
Spin-coated films were produced by dynamic spin-coating on an Ossila spin-coater: a 30 \unit{\micro\litre} dispersion of P3HT nanowires was deposited using a glass syringe onto a substrate (glass slide or silicon wafer) spinning at 1000 revolutions per minute (rpm) and left spinning for three minutes. Spin-coated films were then kept under vacuum for at least half an hour before electronic measurements.

\threesubsection{Atomic Force Microscopy (AFM)}
Images were taken on an Agilent 5500 Atomic Force Microscope (AFM) using force modulation silicon tips (Budget Sensors, 10 \unit{\nm} radius, 3 \unit{\newton\per\meter} force constant, 75 \unit{\hertz} fundamental frequency). Images were recorded in tapping mode at the first harmonic of the tip ($\approx$ 470 \unit{\hertz}).

\threesubsection{Orientation analysis} 
Orientation analysis of P3HT nanowire films (Figures \ref{fig:2}f,g, \ref{fig:S_AFM_droplet}b and \ref{fig:S_AFM_spin-coating}b) was performed using a custom {\sc Matlab} script based on the open-source software OrientationJ\cite{fonck2009effect,rezakhaniha2012experimental,puspoki2016transforms}. Briefly (full description in Ref. \cite{jahne1993spatio}), the local structure of a vector ${\bf x}\equiv(x_1, x_2)$ in 2D space is given by the tensor ${\bf J} = |{\bf x}|^{-1}{\bf x x}^T$. For a greyscale image, a vector space can be created using the partial derivative of the grey values in each dimension $p \in \{1,2\}$ and $q \in \{1,2\}$, so that $ {\bf J} = \begin{bmatrix} J_{1,1} & J_{1,2} \\ J_{2,1} & J_{2,2} \end{bmatrix}$
with components $J_{pq}({\bf x}) = \int_{-\infty}^\infty{\rm d}x'_1{\rm d}x'_2 \,\, w({\bf x}-{\bf x'})\left(\frac{\partial g(x'_1,x'_2)}{\partial x'_p} \frac{\partial g(x'_1,x'_2)}{\partial x'_q} \right)$, being $w$ a window function centerd on $\bf x$. The orientation $\varphi$ of each pixel sharing the end position of vector $\bf x$ can then be determined as $\varphi = \frac{1}{2}\arctan \left( \frac{2J_{1,2}({\bf x})}{J_{2,2}({\bf x})-J_{1,1}({\bf x})} \right)$ (Figure \ref{fig:S_Orientation}). In all our analysis, $w$ was a normalised 2D Gaussian with a standard deviation of 4 pixels, cropped to an area of 11$\times$11 pixels to decrease computational load. 

\threesubsection{{Circular Statistics}} To measure the dispersion of $N$ orientations (defined within periodic boundaries, $\varphi \in (-\frac{\pi}{2},\frac{\pi}{2}]$) in the output matrix from the P3HT nanowire \emph{orientation analysis} (Figures \ref{fig:2}h-i, \ref{fig:S_AFM_droplet}b and \ref{fig:S_AFM_spin-coating}b), the orientations must first be mapped to a full unit circle, i.e. $2\varphi \in (-\pi,\pi]$, before using standard directional statistics \cite{fisher1993}. 
The resultant vector 
$R =\sqrt{\bar c^2 + \bar s^2}$,
where 
$\bar c =\frac{1}{N}\sum_{i}^{N} \cos(2\varphi_i)$ 
and 
$\bar s =\frac{1}{N}\sum_{i}^{N} \sin(2\varphi_i)$, takes the value of 0 for a completely random distribution and 1 when all orientations are perfectly aligned. From these, the mean angle is given by
$\bar\varphi=\frac{1}{2}\atantwo(\frac{\bar s}{\bar c})$ and the standard deviation by 
$\sigma_{\rm s} = \frac{1}{2}\sqrt{-2\ln R}$, 
where the $\frac{1}{2}$ factor is used to map back on the half-unit circle.

\threesubsection{Electronic Characterization}
OFET output and transfer curves were measured in air using an Everbeing probe station connected to a Keithley 2612B System SourceMeter controlled via a custom built {\sc Matlab} program.
Hole mobility $\mu$ (Figure \ref{fig:3}f) was then extracted from transfer curves as \cite{lee2011electrical}
$$\mu(V_{\rm G}) =\frac{1}{C_{\rm SiO_2}}\frac{W}{L}\frac{\partial I_{\rm DS}}{\partial V_{\rm G}}\frac{1}{V_{\rm DS}}$$
where $I_{\rm DS}$, $V_{\rm DS}$ and $V_{\rm G}$ are drain-source current, drain-source voltage and gate voltage, $W$ and $L$ are the inter-electrode width (10 \unit{\um}) and length (1 mm), and $C_{\rm SiO_2} = \frac{\epsilon_r \epsilon_{\rm 0}}{s}$ is the dielectric capacitance per unit area (here, of the underlying ${\rm SiO}_2$ layer, with $s = 200$ nm its thickness, $\epsilon_{\rm r} = 3.9$ its relative permittivity, and $\epsilon_0$ the permittivity of free space). 

\threesubsection{Fabrication of flexible electrodes}
Flexible electrodes (Figures \ref{fig:4} and \ref{fig:flexibleelectodes}) were patterned on 200 \unit{\um}-thick polyethylene terephthalate (PET) substrates coated with conductive indium tin oxide (ITO). A 700 \unit{\nm} dielectric layer of parylene-N was deposited via chemical vapor deposition using an SCS Labcoter 2 (Parylene dimer sublimated at 150 \unit{\degreeCelsius} and then pyrolyzed at 650 \unit{\degreeCelsius}, substrate maintained at room temperature, coating time 3 h). S1818 was spin-coated on top of the substrates (500 rpm for 2 \unit{\s}, then 4000 rpm for 40 \unit{\s}) as a photoresist, which were then soft-baked on a hot plate for 1 min at 115 \unit{\degreeCelsius}. The photoresist was patterned using a Direct Write Laser (DWL 66+, Heidelberg Instruments, 65 mW at 80\% intensity, 50\% attenuation filter, two exposures). The photoresist was then developed in MF-319 for 60 \unit{\s}. 
The patterned substrates were coated in an $\approx$10 \unit{\nm}-thick titanium layer followed by an $\approx$100 \unit{\nm}-thick gold layer via e-beam evaporation. After this step, the photoresist was removed by leaving the patterned substrates in acetone for 5 minutes, followed by placing them in the hot spot of a sonicator for 1 \unit{\s} three times. Before use, the patterned substrates were washed by rinsing with acetone three times, with isopropyl alcohol once, with 50:50 v/v\% isopropyl alcohol/DI water once and finally with DI water three times.

%\medskip
%\textbf{Supporting Information} \par

\medskip
\noindent\textbf{Competing interest} \par
The authors have no competing interests to declare.

\medskip
\noindent\textbf{Author contributions} \par 

Authors contributed as follows, according to CRediT: {\bf RM}: Data Curation, Formal Analysis, Investigation, Methodology, Software, Validation, Visualization, Writing - Original Draft, Writing - Review and Editing. {\bf AR}: Investigation, Methodology, Writing - Original Draft, Writing - Review and Editing. {\bf LMC}: Investigation, Writing - Review and Editing. {\bf PAG}: Investigation, Writing - Review and Editing. {\bf MAP}: Methodology, Writing - Review and Editing. {\bf MC}: Investigation, Writing - Review and Editing. {\bf YC}: Supervision, Writing - Review and Editing. {\bf IP}: Resources, Funding Acquisition, Supervision, Writing - Review and Editing. {\bf MOB}: Conceptualization, Formal Analysis, Resources, Funding Acquisition, Supervision, Writing - Original Draft, Writing - Review and Editing. {\bf BSC}: Conceptualization, Formal Analysis, Resources, Funding Acquisition, Supervision, Writing - Original Draft, Writing - Review and Editing.  {\bf GV}: Conceptualization, Formal Analysis, Resources, Funding Acquisition, Supervision, Project Management, Writing - Original Draft, Writing - Review and Editing.

\medskip
\noindent\textbf{Acknowledgments} \par 
RM, LMC, MOB, BCS and GV gratefully acknowledge the Engineering and Physical Sciences Research Council for supporting this work grant numbers EP/W005875/1, EP/W026813/1. BCS acknowledges funding from the UK Research and Innovation for Future Leaders Fellowship MR/Y003802/1. AR and IP would like to acknowledge funding from the Nikon-UCL Prosperity Partnership on Next-Generation X-Ray Imaging, EPSRC grant EP/T005408/1. We would also like to thank Prof. Khellil Sefiane and Dr. Daniel Orejon Mantecon from the School of Engineering at the University of Edinburgh for initial training and guidance on thermography and Prof. Paul Beard and Dr. Edward Zhang from UCL’s Department of Medical Physics \& Biomedical Engineering for their assistance with parylene coating.

\medskip

\bibliographystyle{MSP}
\bibliography{bib.bib}

\pagebreak
\markboth{}{SUPPLEMENTARY MATERIALS}
\title{Supplementary Materials}
\maketitle

\renewcommand{\thefigure}{S\arabic{figure}}
\renewcommand{\thetable}{S\arabic{table}}
\renewcommand{\theequation}{S\arabic{equation}}
\renewcommand{\thepage}{S\arabic{page}}
\renewcommand{\themovie}{S\arabic{movie}}
\setcounter{figure}{0}
\setcounter{table}{0}
\setcounter{equation}{0}
\setcounter{page}{1} 

\section*{Supplementary Figures}

\begin{figure}[h]
	\centering
	\includegraphics[width=1\textwidth]{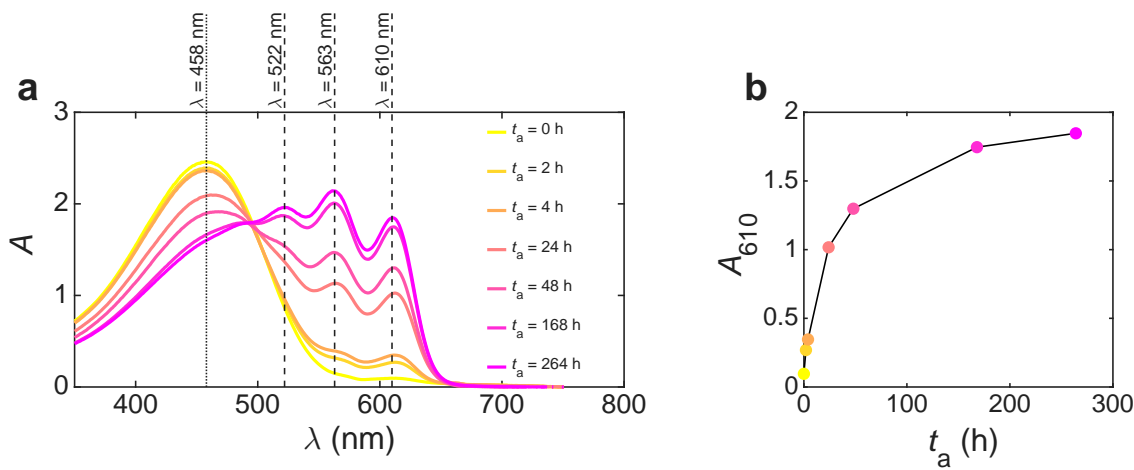}
	\caption{
    \textbf{Growth and aging of P3HT nanowires.}
    (\textbf{a}) UV-Vis absorption spectra (absorbance, $A$) of a 1 mg g\textsuperscript{-1} P3HT solution in 20 wt\% D4/PhCl while aging over 11 days ($t_{\rm a}$ = 264 hours). Absorption of free molecular P3HT ($\lambda$ = 458 nm, dotted line) is seen to decrease relative to its aggregate form ($\lambda$ = 522, 563, 610 nm, dashed lines).  
    Peak positions are similar to previous reports for P3HT nanowires in dichloromethane \cite{kim2011high}. 
    (\textbf{b}) The intensity $A_{610}$ of the absorption peak at 610 nm grows in time starting to saturate after one week, indicating end of nanowire formation.
        }
	\label{fig:S_UV}
\end{figure}

\begin{figure}[h]
	\centering
	\includegraphics[width=1\textwidth]{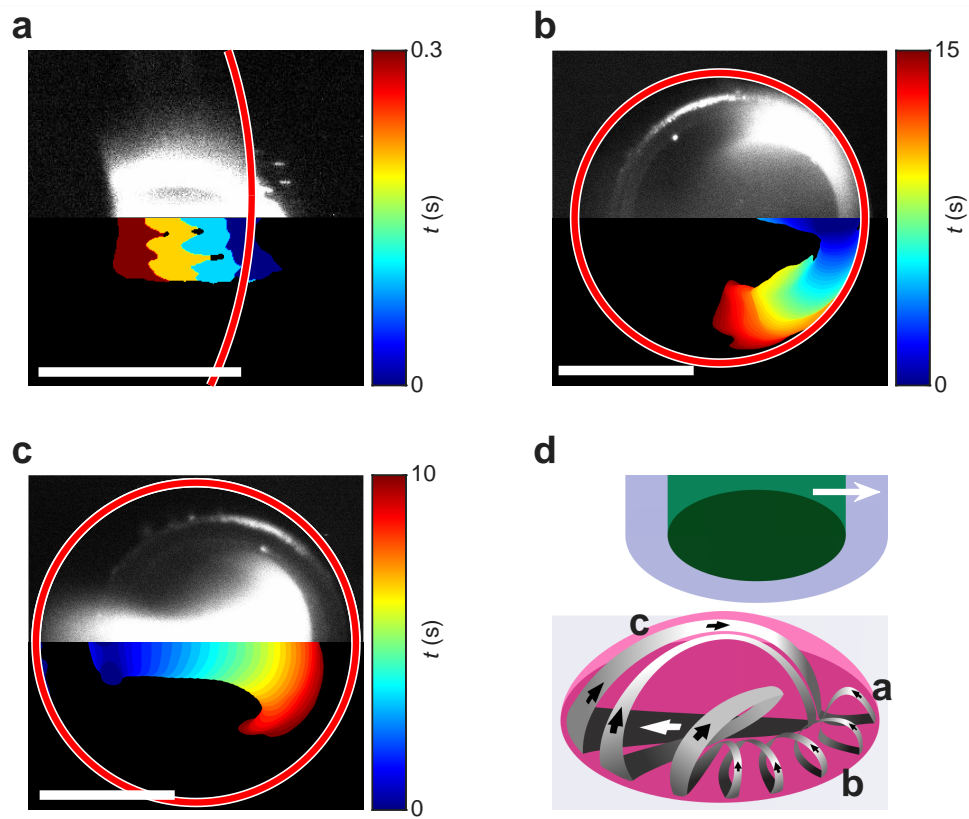}
	\caption{
    \textbf{Extrapolation of flows within a vapor-guided droplet.} 
    (\textbf{a}-\textbf{c}) Exemplary fluorescence images of vapor-guided droplets after being fed a fluorescent dye (top half of images) and corresponding time evolutions of dye transport within the droplets over time (bottom half of images) (Experimental Section). The red circles show the edge of the droplets. (\textbf{a}-\textbf{b}) Feeding the dye at the front shows (\textbf{a}) a rapid vortex forming that (\textbf{b}) spirals towards the side of the droplet before circling towards its center (Fig. 2a, Movie \ref{Movie:FlowFront}). 
    (\textbf{c}) Feeding the dye from the rear of the droplet instead reveals a large central flow from the droplet's back towards its front (Movie \ref{Movie:FlowBack}). 
    Fluorescence images taken at $t$ = 0.3, 15 and 10 \unit{s}, respectively. In \textbf{a}, some dye has not entered the droplet yet and can be seen on the outside of the contact line. Scale bars:  0.5 \unit{mm} in \textbf{a}; 1 \unit{mm} in \textbf{b} and \textbf{c}.
    ({\bf d}) Schematic representation of flows within a vapor-guided droplet as in Fig. 2, highlighting the extrapolated flow lines corresponding to the images in {\bf a}-{\bf c}. 
    }
	\label{fig:flows}
\end{figure}

\begin{figure}[h]
	\centering
	\includegraphics[width=1\textwidth]{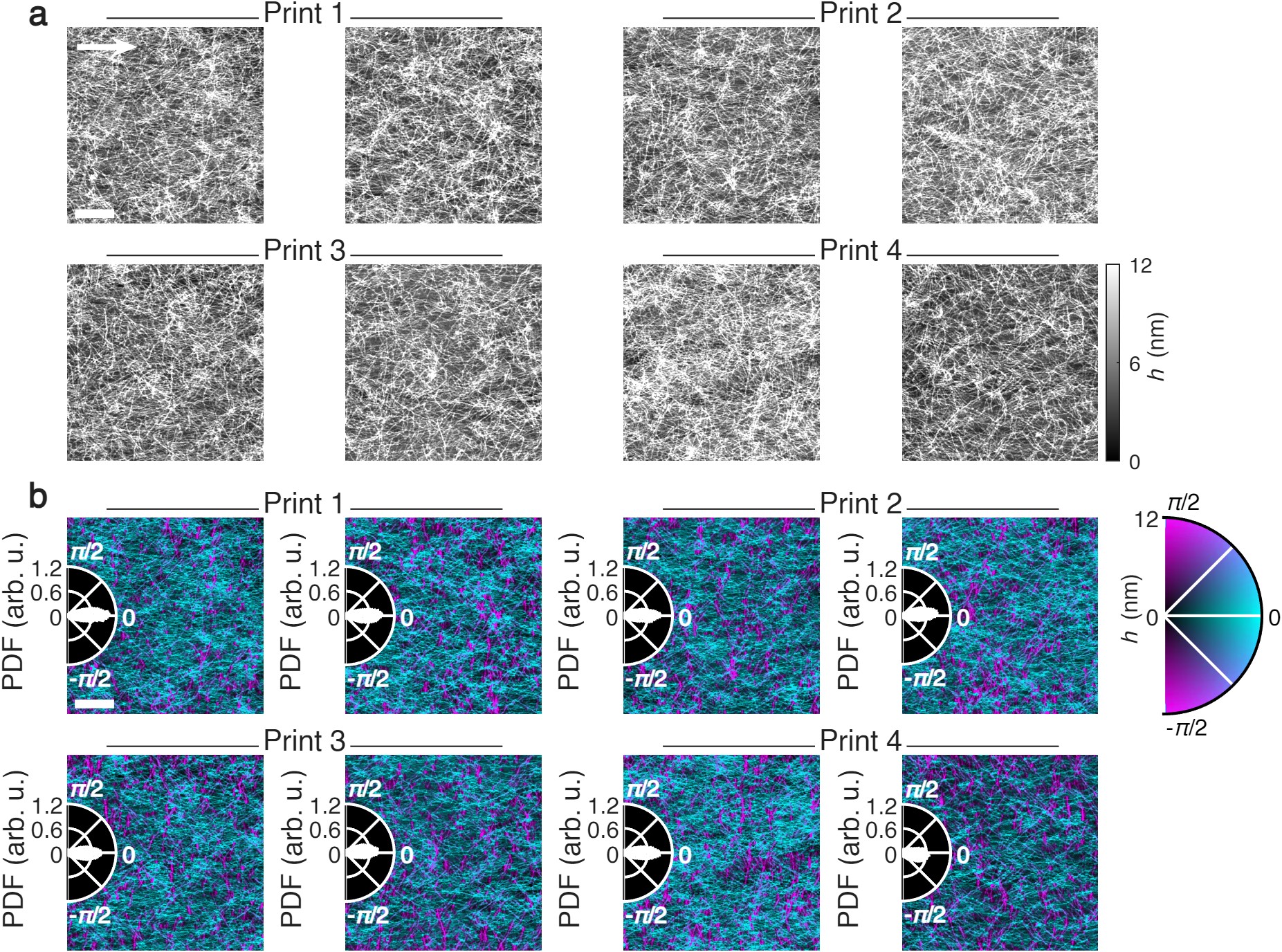}
	\caption{\textbf{Orientation of P3HT nanowire films by droplet printing.}  
    (\textbf{a}) AFM images of P3HT nanowire films printed with vapor-guided droplets on glass slides. The images were taken from two distinct positions along the center of four different prints. Droplet motion was from left to right (white arrow). Scale bar: 1 \unit{\um}.
    (\textbf{b}) Orientation analysis of the images in \textbf{a} (Experimental Section). 
    The colour scale indicates nanowires' alignment in the direction of droplet motion (cyan, 0 rad) and perpendicular to it (magenta, $\pm \frac{\pi}{2}$ rad), with colour intensity showing the AFM measured height $h$. The inset polar plots show the probability distribution function (PDF, white area) of the orientation for the respective images, peaking at 0 rad.
    }
	\label{fig:S_AFM_droplet} 
\end{figure}

\begin{figure}[h]
	\centering
    \includegraphics[width=1\textwidth]{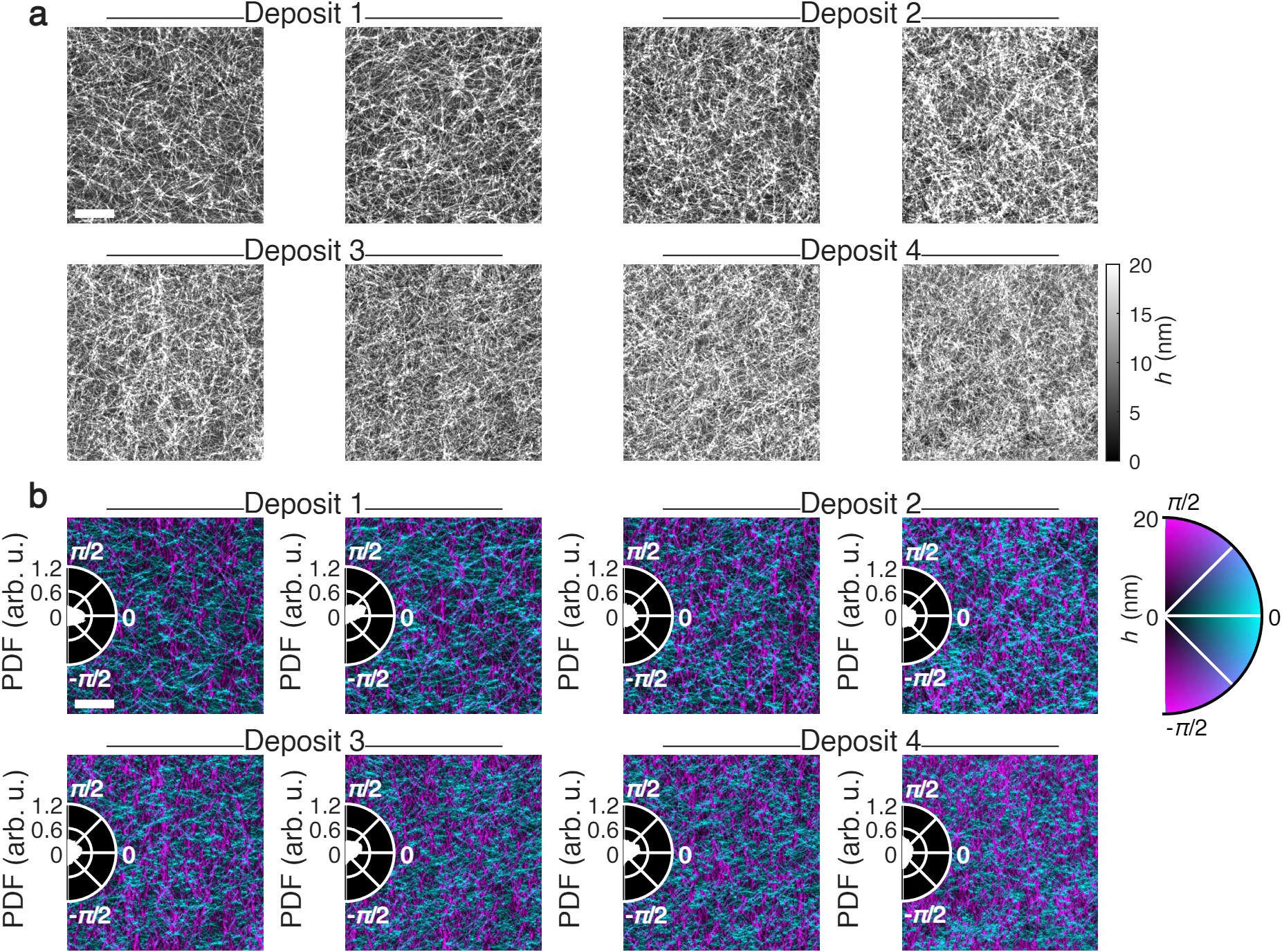}
	\caption{\textbf{Orientation of P3HT nanowire films by spin-coating.}
    (\textbf{a}) AFM images of P3HT nanowire films deposited by spin-coating on glass slides. The images were taken from two distinct positions of four different spin-coated samples. Scale bar: 1 \unit{\um}.
    (\textbf{b}) Orientation analysis of the images in \textbf{a} (Experimental Section). The colour scale indicates nanowires' alignment in the direction of droplet motion (cyan, 0 rad) and perpendicular to it (magenta, $\pm \frac{\pi}{2}$ rad), with colour intensity showing the AFM measured height $h$. The inset polar plots show the probability distribution function (PDF, white area) of the orientation for the respective images, showing more uniform orientation distributions than those in Fig. \ref{fig:S_AFM_droplet}. 
    }
	\label{fig:S_AFM_spin-coating} 
\end{figure}

\begin{figure}[h]
	\centering
	\includegraphics[width=1\textwidth]{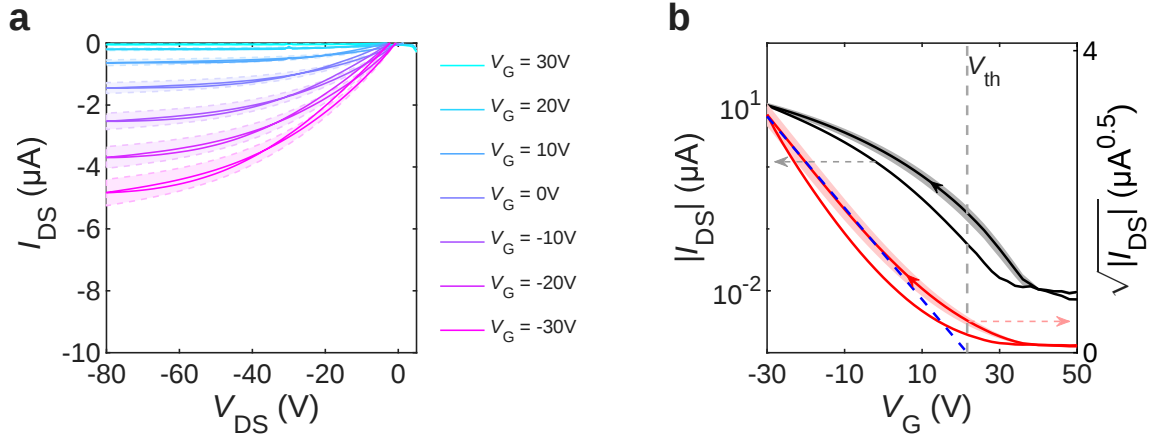}
	\caption{\textbf{OFET performance in the $\parallel$ configuration.} 
    (\textbf{a}) Average OFET output curves (out of eight measurements per curve) showing drain-source current $I_{\rm DS}$ against applied voltage $V_{\rm DS}$ at different gate voltages $V_{\rm G}$ for P3HT nanowire films printed with vapor-guided droplets, measured in the $\parallel$ electrode configuration in air (Fig. 3a).
    (\textbf{b}) Corresponding transfer curves showing $|I_{\rm DS}|$ (black) and $\sqrt{|I_{\rm DS}|}$ (red) as a function of $V_{\rm G}$ at $V_{\rm DS}$ = -80 \unit{\volt}. Threshold voltage $V_{\rm th} = \qty{21.6}{\volt}$ (dashed vertical line), extrapolated from fitting the linear region of the $\sqrt{|I_{\rm DS}|}$ curve (dashed blue line). Dashed arrows point to the vertical axis associated to each curve. Shaded areas show standard error on the forward scan (arrow heads in {\bf b}); backward scan errors are comparable but omitted for clarity.
    }
	\label{fig:S_parallel} 
\end{figure}

\begin{figure}[h]
	\centering
	\includegraphics[width=1\textwidth]{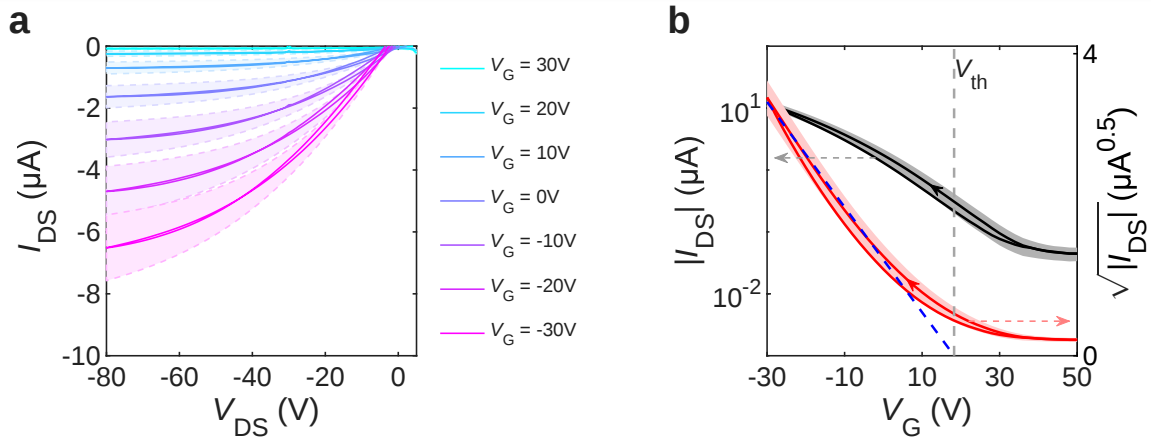}
	\caption{\textbf{OFET performance in spin-coated devices.}
    (\textbf{a}) Average OFET output curves (out of eight measurements per curve) showing drain-source current $I_{\rm DS}$ against applied voltage $V_{\rm DS}$ at different gate voltages $V_{\rm G}$ for P3HT nanowire films deposited by spin-coating and measured in air. 
    (\textbf{b}) Corresponding transfer curves showing $|I_{\rm DS}|$ (black) and $\sqrt{|I_{\rm DS}|}$ (red) as a function of $V_{\rm G}$ at $V_{\rm DS}$ = -80 \unit{\volt}. Threshold voltage $V_{\rm th} = \qty{18.3}{\volt}$ (dashed vertical line), extrapolated from fitting the linear region of the $\sqrt{|I_{\rm DS}|}$ curve (dashed blue line). 
    Dashed arrows point to the vertical axis associated to each curve. Shaded areas show standard error on the forward scan (arrow heads in {\bf b}); backward scan errors are comparable but omitted for clarity.
    }
	\label{fig:S_SC} 
\end{figure}

\FloatBarrier

\begin{figure}[h]
	\centering
	\includegraphics[width=16cm]{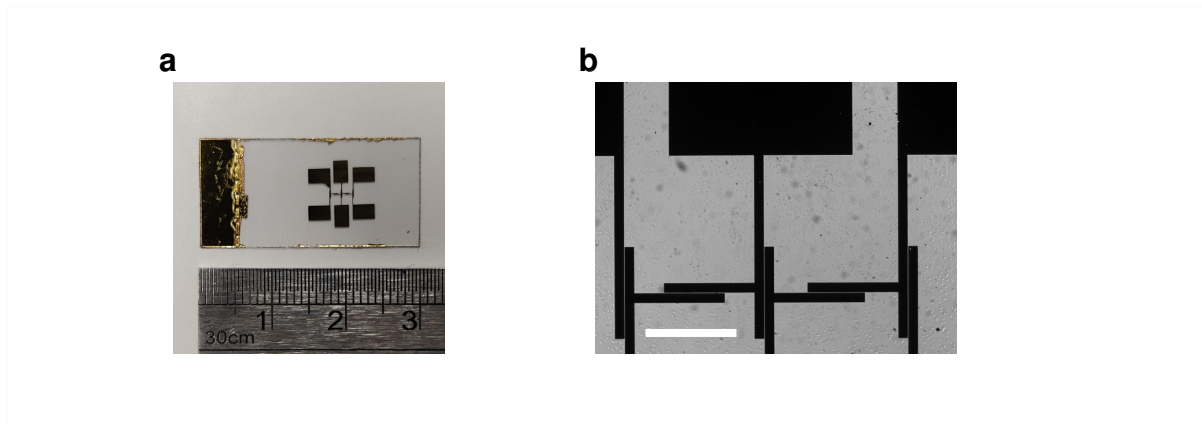}
	\caption{\textbf{Gold electrodes on flexible substrates.}
    (\textbf{a}) Photograph from the top of gold electrodes patterned on PET-ITO-P. Ruler included for scale.
    (\textbf{b}) Bright-field microscopy image showing a close-up of the electrode configuration resembling that in Fig. 3a-b. Scale bar: 1 mm.
    }
	\label{fig:flexibleelectodes} 
\end{figure}

\begin{figure}[h]
	\centering
	\includegraphics[width=1\textwidth]{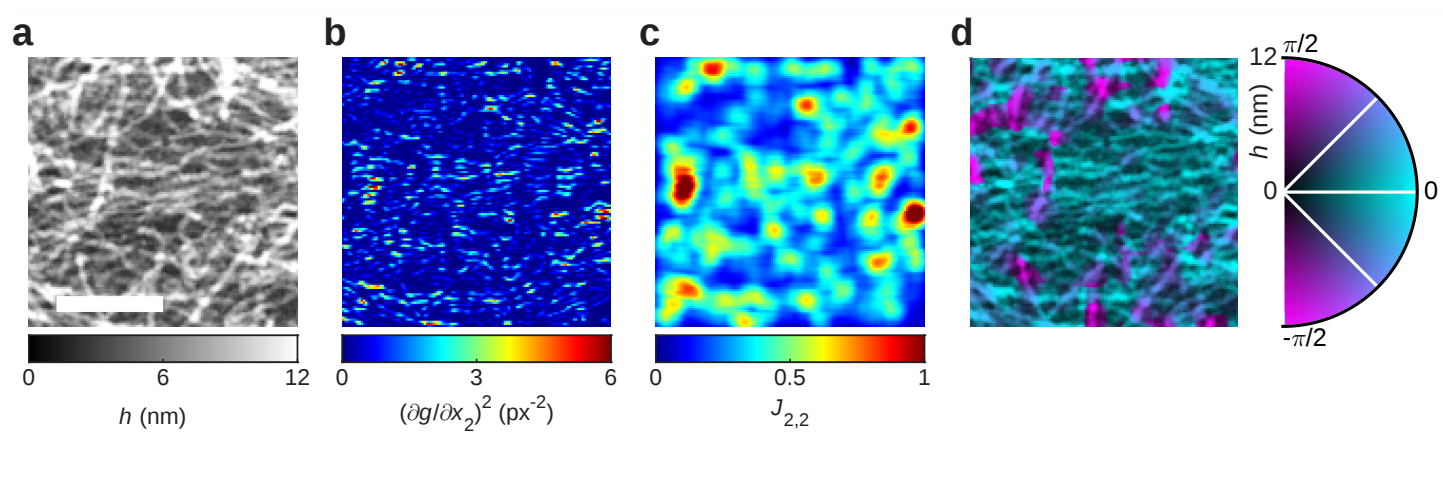}
	\caption{\textbf{Extraction of P3HT nanowire orientation.} (\textbf{a}) A grayscale AFM  image is analyzed to sequentially calculate (\textbf{b}) its gradient functions  $\big(\frac{\partial g(x_1,x_2)}{\partial x_1}\big)^2$, $\big( \frac{\partial g(x_1,x_2)}{\partial x_2}\big)^2$ (shown as example here) and $\frac{\partial g(x_1,x_2)}{\partial x_1}\frac{\partial g(x_1,x_2)}{\partial x_2}$ and (\textbf{c}) the components (here $J_{2,2}$ shown as example) of the local structure tensor $\bf J$, from which (\textbf{d}) the orientation of the P3HT nanowires in the deposited film can be reconstructed (Experimental Section). Scale bar: 500 \unit{\nm}.
    }
	\label{fig:S_Orientation}
\end{figure}

\pagebreak

\section*{Supplementary Movies}

\begin{movie}[h]
	\centering
	\includegraphics[width=0.5\textwidth]{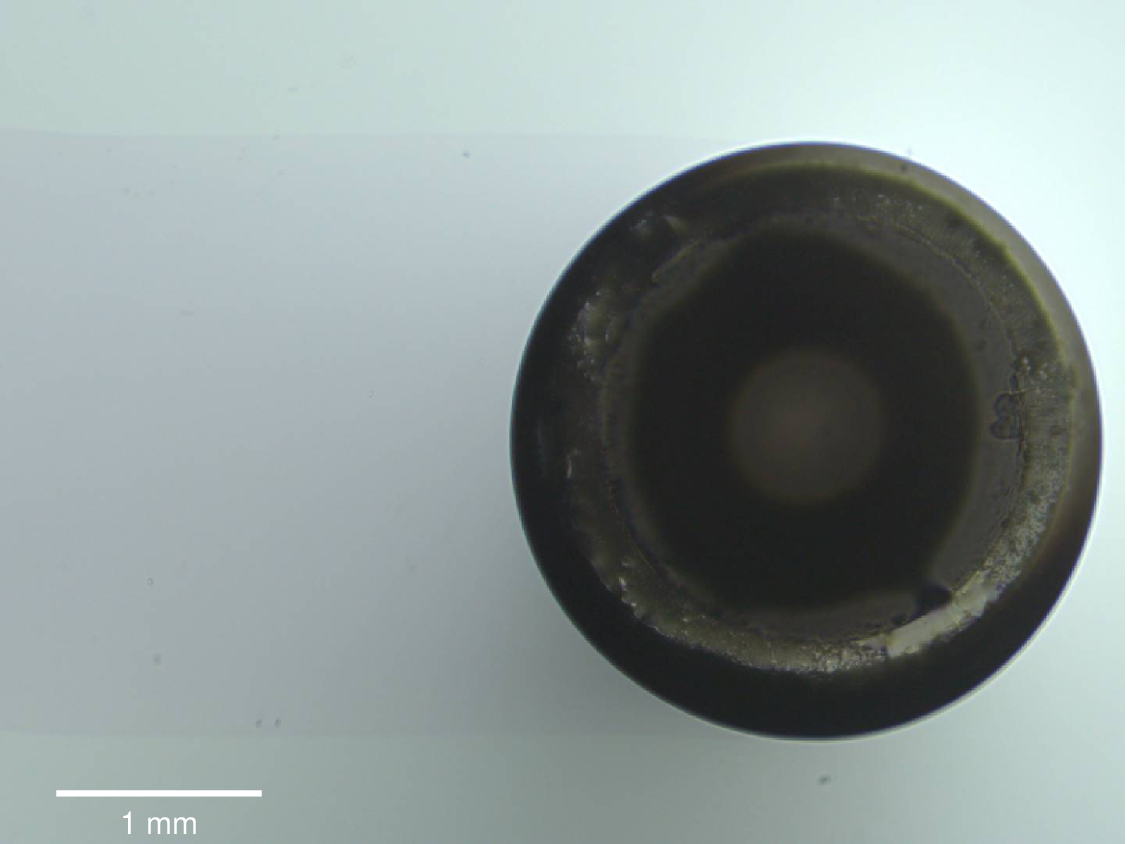}
	\caption{{\bf Printing P3HT nanowire films by vapor-guided droplets.} Printing of a P3HT nanowire film by a vapour-guided droplet (volume $V_{\rm d} = \qty{300}{\nano\litre}$, 20 wt\% D4/PhCl, 1 mg g\textsuperscript{-1} P3HT) moving at a speed of 250 \unit{\um\per\second} onto a glass slide. A thin purple film is visible after the droplet. Droplet moves left to right.
    }
	\label{Movie:Motion} 
\end{movie}

\begin{movie}[h]
	\centering
	\includegraphics[width=0.5\textwidth]{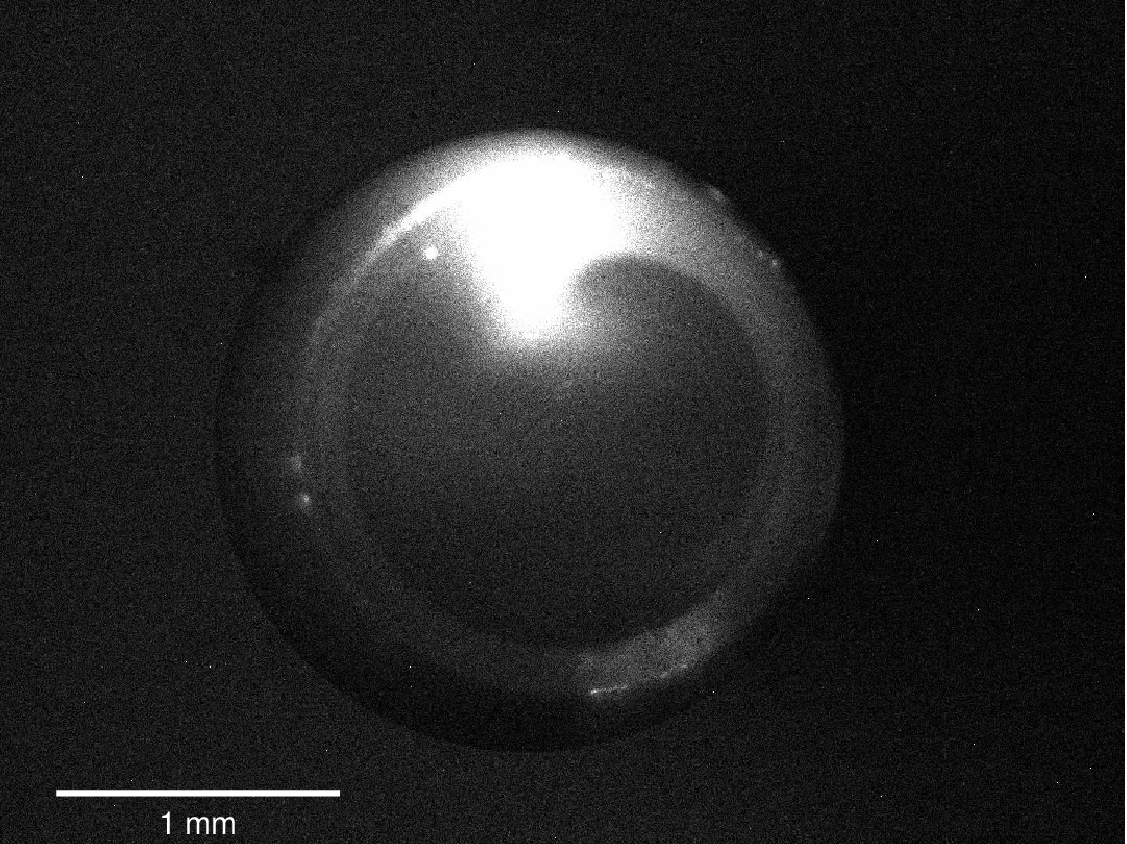}
	\caption{{\bf Flow visualization from the front within a vapor-guided droplet.} A droplet (volume $V_{\rm d} = \qty{300}{\nano\litre}$, 20 wt\% D4/PhCl, 1 mg g\textsuperscript{-1} P3HT) moving at 250 \unit{\um\per\second} over a deposit of fluorescent dye on a glass slide. Droplet moves left to right.
    }
	\label{Movie:FlowFront} 
\end{movie}

\begin{movie}[h]
	\centering
	\includegraphics[width=0.5\textwidth]{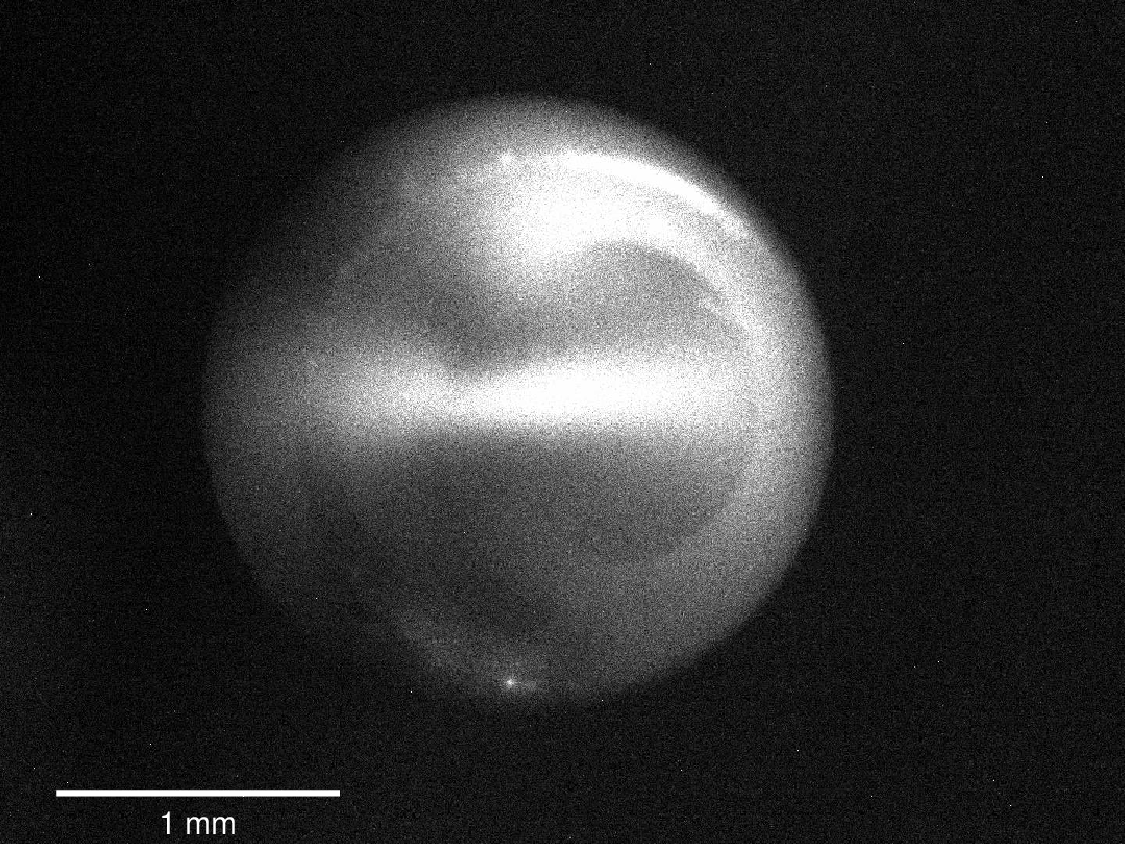}
	\caption{{\bf Flow visualization from the back within a vapor-guided droplet.} 
    A droplet (volume $V_{\rm d} = \qty{300}{\nano\litre}$, 20 wt\% D4/PhCl, 1 mg g\textsuperscript{-1} P3HT) moving (right to left) over a deposit of fluorescent dye on a glass slide, reverting its direction and continuing left to right at 250 \unit{\um\per\second}. Feeding the dye at the opposing side of the droplet than in Supplementary Movie \ref{Movie:FlowFront} allows the visualization of different flow lines within the droplet.
    }
	\label{Movie:FlowBack} 
\end{movie}

\begin{movie}[h]
	\centering
	\includegraphics[width=0.5\textwidth]{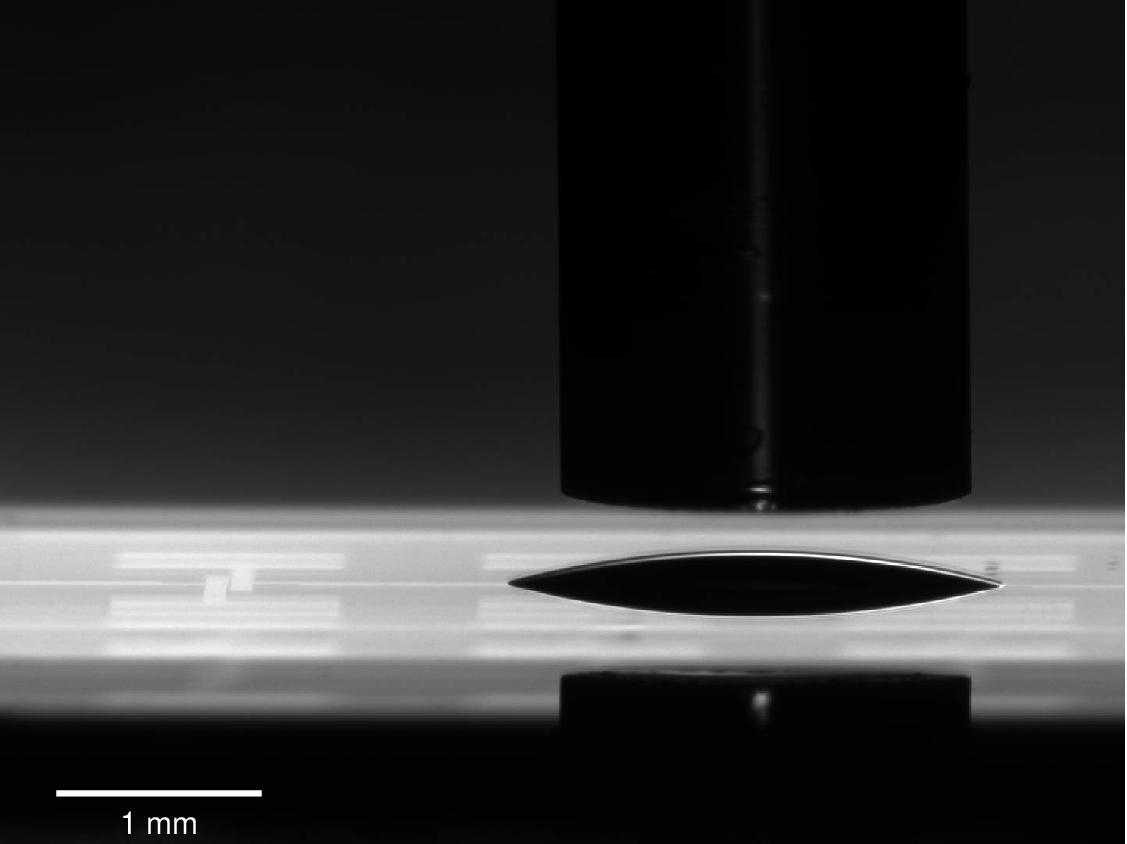}
	\caption{{\bf Printing OFETs by vapor-guided droplets.} Printing of P3HT nanowire films by a vapour-guided droplet (volume $V_{\rm d} = \qty{300}{\nano\litre}$, 20 wt\% D4/PhCl, 1 mg g\textsuperscript{-1} P3HT) moving at a speed of 250 \unit{\um\per\second} onto a oxidised silicon wafer with patterned gold electrodes. 
    The droplet moves smoothly left to right over the height steps ($\approx$45 \unit{\nm}) provided by the electrodes without pinning. 
    }
	\label{Movie:MotionSide} 
\end{movie}

\begin{movie}[h]
	\centering
	\includegraphics[width=0.5\textwidth]{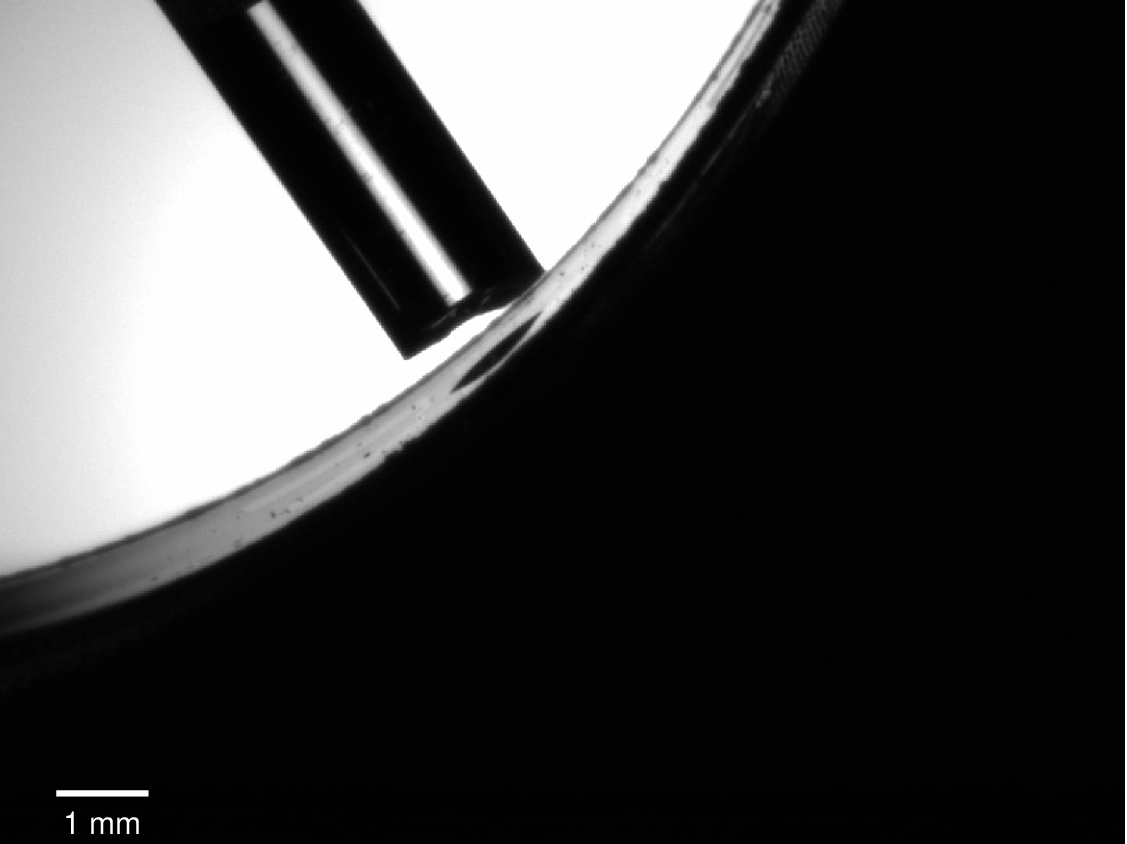}
	\caption{\textbf{Vapor-guided droplet moving up a curved substrate.} A vapor-guided droplet (volume $V_{\rm d} = \qty{100}{\nano\litre}$, 20 wt\% D4/PhCl, 1 mg g\textsuperscript{-1} P3HT) moving up a curved flexible substrate patterned with gold electrodes with a tangential velocity of 250 \unit{\um\per\second}. Droplet moves left to right.
    }
	\label{Movie:Uphill} 
\end{movie}

\FloatBarrier
%\bibliographystyle{MSP}
%\bibliography{bib.bib}
\end{document}